\documentclass{jfm}
\usepackage[longnamesfirst]{natbib}
\usepackage[dvips]{graphicx}
\usepackage{amsmath}
\usepackage{amsfonts}
\usepackage{amssymb}
\usepackage{upmath}
\usepackage{color}
\usepackage{psfrag}
\usepackage{pstricks}
\usepackage{labelfig}
\usepackage{rotating}
\usepackage{stmaryrd}
\usepackage{framed}

\newcommand{\bsym}[1]{\boldsymbol{#1}}

\newcommand{\tensor}[1]{\ensuremath{\mathsfbi{#1}}}

\newcommand{\textfrac}[2]{\textstyle{\frac{#1}{#2}}}

\newcommand{\dd}{\ensuremath{\,\mathrm{d}}}
\newcommand{\pd}{\upartial}

\newcommand{\e}{\mathrm{e}}
\newcommand{\ci}{\mathrm{i}}
\newcommand{\Real}[1]{\mbox{Re}\left(#1\right)}

\newcommand{\fd}[2]{\frac{\dd #1}{\dd #2}}
\newcommand{\fdd}[2]{\frac{\dd^2 #1}{\dd #2^2}}

\newcommand{\fpd}[2]{\ensuremath{\frac{\pd #1}{\pd #2}}}

\renewcommand{\bar}[1]{\overline{#1}}

\renewcommand{\ge}{\geqslant}
\renewcommand{\le}{\leqslant}

\newcommand{\Ri}{\ensuremath{\mbox{\textit{Ri}}}}

\shortcites{Sparks1997,Plourde2008,Odier2009,Scase2006a,Woodhouse2013}

\title[Unsteady turbulent buoyant plumes]{Unsteady turbulent buoyant plumes}
\author[M.J. Woodhouse, J.C. Phillips \& A.J. Hogg]
{M.\ns J.\ns W\ls O\ls O\ls D\ls H\ls O\ls U\ls S\ls E$^{1,2}$,\ns J.\ns C.\ns P\ls H\ls I\ls L\ls L\ls I\ls P\ls S$^{2}$,\ns \and\ls A.\ns J.\ns H\ls O\ls G\ls G$^{1}$
\affiliation{$^{1}$School of Mathematics, University of Bristol, University Walk, Bristol, BS8 1TW, UK \\[\affilskip]
$^{2}$School of Earth Science, University of Bristol, Wills Memorial Building, Queen's Road, Bristol, BS8 1RJ, UK
}
}

\begin{document}
\maketitle

\begin{abstract}
We model the unsteady evolution of turbulent buoyant plumes following temporal changes to the source conditions.  The integral model is derived from radial integration of the governing equations expressing the conservation of mass, axial momentum and buoyancy in the plume.  The non-uniform radial profiles of the axial velocity and density deficit in the plume are explicitly described by shape factors in the integral equations; the commonly-assumed top-hat profiles lead to shape factors equal to unity. The resultant model for unsteady plumes is hyperbolic when the momentum shape factor, determined from the radial profile of the mean axial velocity in the plume, differs from unity.  The solutions of the model when source conditions are maintained at constant values are shown to retain the form of the well-established steady plume solutions.  We demonstrate through a linear stability analysis of these steady solutions that the inclusion of a momentum shape factor in the governing equations that differs from unity leads to a well-posed integral model.  Therefore, our model does not exhibit the mathematical pathologies that appear in previously proposed unsteady integral models of turbulent plumes.  A stability threshold for the value of the shape factor is also identified, resulting in a range of its values where the amplitude of small perturbations to the steady solutions decay with distance from the source.  The hyperbolic character of the system of equations allows the formation of discontinuities in the fields describing the plume properties during the unsteady evolution, and we compute numerical solutions to illustrate the transient development of a plume following an abrupt change in the source conditions.  The adjustment of the plume to the new source conditions occurs through the propagation of a pulse of fluid through the plume.  The dynamics of this pulse are described by a similarity solution and, through the construction of this new similarity solution, we identify three regimes in which the evolution of the transient pulse following adjustment of the source qualitatively differ.
\end{abstract}

\section{Introduction}

Turbulent buoyant plumes occur in numerous industrial and environmental settings \citep{Woods2010}.  Industrial examples include ventilation and heating \citep[e.g.][]{Baines1969,Linden1990}, industrial chimneys \citep[e.g.][]{Slawson1967} and waste-water disposal \citep[e.g.][]{Koh1975}.  In the natural environment, turbulent plumes are found in meteorological \citep[e.g.][]{Emanuel1994,Stevens2005} oceanographical \citep[e.g.][]{Speer1989,Straneo2015}, and volcanological \citep[e.g.][]{Woods1988,Sparks1997,Woodhouse2013} settings.  Models of steady plumes, based on the integral modelling approach pioneered by \citet{Zeldovich1937}, \citet{Schmidt1941}, \citet{Rouse1952}, \citet{PriestleyBall1955} and \citet{Morton1956}, have been applied extensively to understand plume dynamics.  In this approach, the turbulent flow in the plume is described on a time-scale that is longer than the eddy turnover time (the time scale that characterises the turbulent motion), and therefore the complicated turbulent motions are not explicitly modelled and only the evolution of mean flow quantities with distance from the source are described.  A further simplification is obtained by integrating the mean flow quantities over the cross section of the plume, resulting in a system of nonlinear ordinary differential equations that describe the spatial development of the steady integral flow quantities such as the fluxes of mass, momentum and buoyancy.

To obtain the integral fluxes, it is necessary to specify the radial and azimuthal dependence of the three-dimensional mean fields.  When the flow domain and ambient conditions do not introduce an asymmetry, the time-averaged flow quantities are swirl free (i.e. there is no azimuthal dependence of the mean flow quantities).  Experiments suggest that, at sufficiently high Reynolds number and for an unstratified ambient fluid, buoyant plumes attain self-similar radial profiles for both the mean axial velocity and the concentration of the species that gives rise to the density difference between the plume and the ambient fluid sufficiently far from the fluid source \citep{Papanicolaou1988,Shabbir1994,Wang2002,Ezzamel2015}.  Furthermore, higher-order turbulent quantities such as the turbulent intensity (the root mean square of velocity component fluctuations and concentration fluctuations from the mean) and turbulent stresses also evolve with self-similar radial profiles \citep{Papanicolaou1988,Wang2002,Ezzamel2015}.  The cross-sectional integration then results in a system of ordinary differential equations that describes the evolution of the mass flux, momentum flux and buoyancy flux with distance from the source.  For stratified ambient environments, the detailed structure and evolution of the mean and turbulent quantities have been scrutinised less thoroughly.  However, the applicability of integral models assuming similarity of the mean flow profiles has been demonstrated through comparison of model predictions to laboratory experiments \citep[see e.g.][]{Morton1956,List1982} and in field-scale applications \citep[see e.g.][]{Turner1986,Woods1988,Speer1989,Kaye2008}.

While averaging over the turbulent timescale and integrating over the plume cross-section significantly simplifies the mathematical description of the plume dynamics, detailed information about the turbulent flow is not fully captured.  The integral model of \citet{Morton1956} incorporates the turbulent nature of the flow through a parameterization of turbulent mixing as an inflow of fluid from the ambient to the plume, referred to as entrainment.  The velocity of the entraining flow is linearly related to the mean axial velocity scale of the plume (as dimensional analysis demands, since the only velocity scale that remains following time-averaging and cross-sectional integration is the mean axial velocity of the plume).  \citet{Morton1956} take a constant entrainment coefficient, and the application of this model has been successful in describing steady buoyant plumes over a wide range of scales, from the laboratory \citep[see e.g.][]{Morton1956,List1982} to plumes from large volcanic eruptions \citep[e.g.][]{Woods1988}, illustrating the success of the integral modelling approach with a constant entrainment coefficient.

Detailed examination of laboratory experiments of turbulent plumes suggests that the entrainment coefficient does not have a constant universal value, but rather evolves as the flow develops \citep{Wang2002,Kaminski2005,Ezzamel2015}.  In particular, for plumes that are strongly forced with a flux of momentum at the source (referred to as buoyant jets), the entrainment coefficient transitions from a value appropriate for jets to the value for plumes as the flow becomes increasingly driven by buoyancy.  Integral models that include an evolving entrainment coefficient have been proposed \citep[see e.g.][]{Fox1970,Kaminski2005,Carazzo2006,Craske2015a,Craske2015b}, building on the integral modelling approach of \citet{PriestleyBall1955} whereby an integral expression for the conservation of axial kinetic energy is used with conservation of momentum to derive an integral expression for conservation of mass.  (This modelling approach is discussed further in appendix \ref{sec:app energy closure}.)  However, the assumption of a constant entrainment coefficient captures the leading order behaviour of turbulent buoyant plumes over a wide range of scales \citep{Turner1986}, and for fully-developed turbulent plumes sufficiently far from the source, a constant entrainment coefficient is appropriate and represents the similarity of the flow profiles and the turbulent entrainment processes at different heights in the plumes \citep{Turner1986,Ezzamel2015}.

The applicability of steady models to describe the inherently unsteady, turbulent motion relies on a separation of time scales.  If the conditions at the plume source and in the ambient are held steady, the steady integral models well describe the plume behaviour on time scales that are long compared to the eddy turn-over time \citep{Woods2010}.  However, if the source conditions change on a time scale that is longer than the turbulent fluctuations, then a signature of the source variation may be seen in time-averaged plume dynamics downstream of the source \citep[e.g.][]{Scase2008,Scase2009}.  Unsteady sources occur frequently in natural settings, for example as the strength of a volcanic source changes in magnitude during an eruption.  In addition, temporal changes in ambient conditions on a time scale similar to the ascent time of a fluid parcel are likely to result in a transient response of the plume.

To model unsteady plumes, integrals models have been proposed \citep{Delichatsios1979,Yu1990,Vulfson2001,Scase2006a,Scase2006b,Scase2008,ScaseAspdenCaulfield2009,Scase2009} that extend the modelling approach of \citet{Morton1956} while retaining some of the underlying assumptions of the steady model.  In particular, the unsteady models capture the variations of flow quantities on a time-scale that is longer than the eddy turn-over time, and assume that the radial profiles of the mean axial velocity and the concentration of the buoyancy generating species remain in a self-similar form throughout the evolution.  However, given the difficulty in obtaining robust experimental results for plumes with time-varying source conditions, the underlying assumptions have yet to be scrutinised in detail.  Numerical simulation of the governing equations can be used as a surrogate for laboratory experiment and allow detailed investigations of the turbulent flow properties throughout the modelled domain \citep{Jiang2000,Plourde2008,Craske2013,Craske2015a}.   The physical basis of the unsteady models have been further questioned by \citet{Scase2012} in an analysis the stability of the steady solutions of the models of \citet{Delichatsios1979}, \citet{Yu1990}, and \citet{Scase2006a} to small harmonic perturbations at the source.  \citet{Scase2012} find that the perturbations grow as they propagate through the plume, and that the growth rate increases without bound as the frequency of the source oscillations is increased; therefore the models are ill-posed, suggesting a critical physical process has been neglected \citep{Joseph1990}.

In an attempt to `regularize' the unsteady plume models, \citet{Scase2012} introduce a phenomenological diffusive term into the equation for conservation of axial momentum to represent turbulent mixing processes; it curtails the unbounded growth of short wavelength perturbations thus leading to a well-posed model.  The form of the diffusive term has been investigated by \citet{Craske2015a} for momentum-driven (non-buoyant) jets using direct numerical simulations; the numerical simulations suggest that the diffusive term proposed by \citet{Scase2012} does not describe well the transient evolution of momentum-driven jets.  Here we find that the diffusive term introduced by \citet{Scase2012} leads to a new pathology in the system of equations; the steady states of the `regularized' model are spatially unstable and therefore cannot be realised.

We show here that the ill-posedness in the unsteady models analyzed by \citet{Scase2012} is due to the assumption of a top-hat profile for the mean axial velocity (i.e. the axial velocity at any height  is assumed to be radially invariant within the plume, and zero outside of the plume) and therefore a failure to account for the non-uniform radial profile of the mean axial velocity of the plume.  Non-uniform radial profiles for the axial velocity were examined in the early studies of steady plumes by \citet{PriestleyBall1955} and \citet{Morton1956}.  However, solutions of steady plume models have the same form whether top-hat or non-uniform radial profiles for the mean axial velocity are adopted, as dimensional analysis demands, with only changes to coefficients in the solutions \citep{Morton1956,Linden2000,Kaye2008} and many subsequent analyses of steady plumes have adopted the top-hat formulation.

When a top-hat velocity profile is adopted, the cross-sectionally averaged mass and momentum of the plume are transported with the same rate, but the transport rates differ when the radial profile of the mean axial velocity is non-uniform.  We therefore propose an integral model of unsteady plumes that explicitly accounts for the different transport rates of the cross-sectionally averaged mass and axial momentum of the plume, by introducing a `shape factor' in the equation for the conservation of momentum that differs from unity when non-uniform radial profiles of the mean axial velocity are assumed.  A shape factor that differs from unity in shallow-water hydraulic models has been shown to fundamentally alter solutions of the system of equations due to a change in the characteristics of the hyperbolic system \citep{Hogg2004}.  Here we show that including a shape factor changes the character of the system of equations describing unsteady plumes and leads to a well-posed system of equations without the need to include diffusive terms.

A similar approach to regularising an unsteady integral model of momentum driven jets has been proposed recently by \citet{Craske2015a,Craske2015b}.  Through analysis of their direct numerical simulations, \citet{Craske2015a,Craske2015b} develop a well-posed integral model of unsteady jets that includes a description of dispersion in the plume, resulting from the non-uniform radial profile of the mean axial velocity that leads to different transport rates for mass, axial momentum and kinetic energy \citep[referred to as type I dispersion by ][]{Craske2015a,Craske2015b}), and the deviation of the radial profile of mean axial velocity from a self-similar form (referred to as type II dispersion).  Type II dispersion is important in jets, where the flow structure evolves significantly in the neighbourhood of the source even for temporally invariant source conditions\citep{Wang2002,Kaminski2005,Ezzamel2015}, but is likely less important for fully developed turbulent plumes for which the radial profiles are in self-similar forms \citep{Ezzamel2015}.  However, \citet{Craske2015a,Craske2015b} show that explicitly accounting for the non-uniform radial profile of mean axial velocity (type I dispersion) in the integral equations is critical to the well-posedness of the unsteady integral model for jets.

In this contribution we demonstrate, through an analysis of the temporal evolution of small perturbations to  steady solutions, that an integral model of unsteady plumes is well-posed when the momentum shape factor differs from unity.  Furthermore, we identify a stability threshold in the value of the shape factor above which the amplitude of small perturbations decay.  The system of equations we propose is hyperbolic, with a characteristic structure that, in certain situations, allows for the formation of `shocks' during the transient evolution.  Through the construction of similarity solutions, we identify scaling relationships that describe the propagation and growth of a transient pulse that is advected through the plume following an abrupt change in the source conditions, and determine the regimes in which shocks are formed.

This paper is organized as follows.  In \S\ref{sec:model intro} we provide a derivation of an integral model of unsteady buoyant plumes, beginning from the Reynolds-averaged Navier--Stokes equations for a high Reynolds number flow together with an advection-diffusion equation for the transport of a scalar species the concentration of which determines the local fluid density.  We demonstrate that the use of integral flow quantities, obtained through integration over a plume cross-section, requires the inclusion of shape factors for the transport rates of the (cross-sectionally averaged) axial momentum and buoyancy of the plume.  We show that a momentum shape factor modifies only slightly the classical power-law solutions of \citet{Morton1956}.  In \S\ref{sec:Scase problems} we re-examine the phenomenological diffusive term introduced by \citet{Scase2012} to `regularize' the ill-posed unsteady plume model, and show this model introduces a new pathology into the system of equations whereby steady solutions are spatially unstable.  We therefore examine the mathematical structure and well-posedness of an unsteady integral model for plumes that accounts for the non-uniform radial profile of the mean axial velocity through a momentum shape factor, and we demonstrate in \S\ref{sec:well posed} that this leads to a well-posed system of equations.  Numerical solutions of our unsteady model are presented in \S\ref{sec:numerical solutions} to support the mathematical analysis.  In \S\ref{sec:similarity solns} we consider the evolution of a plume following an abrupt change in the source buoyancy flux and show that the adjustment of the plume occurs through the propagation of a pulse whose dynamics is described by a similarity solution.  Finally, in \S\ref{sec:conclusion} we discuss the implications of our mathematical model and draw our conclusions.

\section{An integral model for unsteady turbulent buoyant plumes}
\label{sec:model intro}

We model an unsteady turbulent buoyant plume formed due to the release of a fluid from a point source into an otherwise quiescent ambient fluid of a different density.  A cylindrical coordinate system is adopted, with $\bsym{\hat{r}}$ and $\bsym{\hat{z}}$ denoting unit vectors in the radial and vertical directions, respectively.  The plume and ambient are composed of incompressible fluids and the plume is assumed to have a circular (time-averaged) cross-section.  The velocity field, $\bsym{u}$, is assumed to be axisymmetric, with $\bsym{u} = u\bsym{\hat{r}} + w\bsym{\hat{z}}$.  We further assume the plume is slender such that $R/H \ll 1$ where $R$ and $H$ denote typical length scales in the radial and vertical direction respectively, that the Reynolds number of the emitted fluid at the source is sufficiently high such that the turbulent motion is fully developed throughout the flow domain, and that the plume fluid transports a scalar species (such as heat or salt) with the plume density linearly related to the concentration of the species.

Turbulence in the plume is responsible for the entrainment of ambient fluid and mixing of the plume and ambient fluids, and its role is central to the ensuing dynamics.  We therefore adopt the Reynolds-averaged Navier--Stokes equations to describe the fluid motion, with a Reynolds-averaged advection-diffusion equation to describe the transport of the scalar species that results in the density difference between the plume and the ambient fluids, taking $u = \bar{u} + u'$, $w = \bar{w} + w'$, and $g_{r} = \bar{g_{r}} + {g_{r}}$, where $\bar{u}$ and $u'$ denote the ensemble average and fluctuation about the average of $u$, respectively, and similarly for $w$ and $g_{r}$, and where $g_{r} = g\left(\rho_{a}-\rho\right)/\rho_{0}$ is the reduced gravity, with $g$ denoting the gravitational acceleration, $\rho$ and $\rho_{a}$ denoting the density of the plume and ambient fluids, respectively, and $\rho_{0}$ is a characteristic density scale.  The equations for the conservation of (the ensemble averaged) mass, axial momentum, and reduced gravity are then
\begin{subeqnarray}
	& & \frac{1}{r}\fpd{}{r}\left(r \bar{u}\right) + \fpd{\bar{w}}{z} = 0, \slabel{eqn:RA mass} \\[3pt]
	& & \fpd{\bar{w}}{t} + \frac{1}{r}\fpd{}{r}\left(r \bar{u} \bar{w}\right) + \fpd{}{z}\left(\bar{w}^{2}\right) = \bar{g_{r}} - \frac{1}{r}\fpd{}{r}\left(r\overline{u'w'}\right) - \fpd{}{z}\left(\overline{{w'}^{2}}\right), \slabel{eqn:RA mom} \\
	& & \fpd{\bar{g_{r}}}{t} + \frac{1}{r}\fpd{}{r}\left(r \bar{g_{r}} \bar{u}\right) + \fpd{}{z}\left(\bar{g_{r}}\bar{w}\right) = \frac{g}{\rho_{0}}\left(\fpd{\rho_{a}}{t} + \fpd{}{z}\left(\bar{w}\rho_{a}\right)\right)  - \frac{1}{r}\fpd{}{r}\left(r \overline{{g_{r}}' u'}\right) - \fpd{}{z}\left(\overline{{g_{r}}' w'}\right). \nonumber \\
	& & {} \slabel{eqn:RA buoy}
	\label{eqn:RA eqns}
\end{subeqnarray}
Note, in \eqref{eqn:RA eqns} we have made use of the plume slenderness and have invoked the Boussinesq approximation whereby differences in density are neglected except where they are multiplied by the gravitational acceleration.  Furthermore, we have neglected molecular diffusion terms under the assumption that the Reynolds number and the Pecl{\'e}t number are sufficiently large so that mixing is dominantly due to turbulence.

Integral equations are obtained by integrating each of equations \eqref{eqn:RA eqns} over a cross-section of the plume.  We define a surface $r = b(z,t)$ representing the boundary of the plume over which entrainment of ambient fluid into the plume occurs and impose the boundary condition on this surface,
\begin{equation}
	\bar{u}(b,z,t) = \fpd{b}{t} + \bar{w}(b,z,t)\fpd{b}{z} + u_{e}, \label{eqn:kinematic bc}
\end{equation}
where the entrainment velocity, $u_{e}$, is the radial velocity of the ambient fluid across $r=b(z,t)$ (note $u_{e}<0$ when ambient fluid is entrained into the plume).  We note that this approach differs from that taken by \citet{Craske2015a,Craske2015b} who adopt the characteristic length and velocity scales defined through the cross-sectionally averaged fluxes of mass and axial momentum which themselves are defined in terms of an additional length scale at which the mean axial velocity can be considered negligible in contrast to the mean axial velocity at the centreline.  As turbulent plumes typically exhibit non-uniform radial profiles of axial velocity and concentration of species, with decaying velocity and concentration as the radial distance from the plume axis, in laboratory experiments or direct numerical simulations there is a choice as to how to define the plume edge, taking, for examples, the radial distance at which the axial velocity or concentration reach a specified threshold, or the radial distance at which the mass flux included in a cross-sectional integral captures a specified proportion of the total mass flux induced by the plume.  We discuss below how these choices are represented in our model.

Integration of the point-wise conservation equations \eqref{eqn:RA eqns} over a plume cross-section, using Leibniz's theorem for interchanging differentiation and integration \citep{Flanders1973} together with the boundary condition \eqref{eqn:kinematic bc}, gives
\begin{subeqnarray}
	& & b\fpd{b}{t} + \fpd{}{z}\int_{0}^{b} r\bar{w}\dd r = -b u_{e}, \slabel{eqn:int mass full} \\[3pt]
	& & \fpd{}{t}\int_{0}^{b} r\bar{w}\dd r + \fpd{}{z}\int_{0}^{b} r\bar{w}^{2}\dd r = \int_{0}^{b} r\bar{g_{r}}\dd r - \fpd{}{z}\int_{0}^{b}r\bar{{w}'^{2}}\dd r \nonumber \\[3pt]
	& & \qquad - b\bar{w}(b,z,t)u_{e} - b\overline{u'w'}(b,z,t) + b\overline{{w'}^{2}}(b,z,t)\fpd{b}{z}, \slabel{eqn:int mom full} \\[3pt]
	& &\fpd{}{t}\int_{0}^{b} r\bar{g_{r}}\dd r + \fpd{}{z}\int_{0}^{b} r\bar{w}\bar{g_{r}}\dd r = \frac{g}{\rho_{0}}\int_{0}^{b} \left(\fpd{\rho_{a}}{t} + \bar{w}\fpd{\rho_{a}}{z}\right) r\dd r \nonumber \\[3pt]
	& & \qquad - \fpd{}{z}\int_{0}^{b} r\overline{g_{r}'w'}\dd r - b\bar{g'_{r}}(b,z,t)u_{e} - b\overline{u'g_{r}'}(b,z,t) - b\overline{w'g_{r}'}(b,z,t)\fpd{b}{z}, \slabel{eqn:int buoy full}
	\label{eqn:full int eqns}
\end{subeqnarray}
representing conservation of mass, momentum and buoyancy, respectively.

We define the integral volume flux as $Q = b^{2}W$ (note, under the Boussinesq assumption, $Q$ can also be referred to as the (specific) mass flux), the (specific) momentum flux as $M = b^{2}W^{2}$ and the buoyancy flux as $F = b^{2}W g'$.  Here $W$ and $G'$ denote the cross-sectionally averaged mean axial velocity and reduced gravity of the plume, respectively, and are given by
\refstepcounter{equation}
$$
	W = \frac{2}{b^{2}}\int_{0}^{b} r \bar{w}\dd r, \quad \text{and} \quad 	G' = \frac{2}{b^{2}}\int_{0}^{b} r\bar{g_{r}}\dd r.
	\eqno{(\theequation{\mathit{a},\mathit{b}})}
	\label{eqn:integral W g'}
$$
We then have
\refstepcounter{equation}
$$
	\int_{0}^{b} r\bar{w}^{2}\dd r = \tfrac{1}{2}SM, \quad \text{and} \quad \int_{0}^{b} r \bar{w}\bar{g_{r}}\dd r = \tfrac{1}{2}\phi F,
	\eqno{(\theequation{\mathit{a},\mathit{b}})}
	\label{eqn:integral w2 wg'}
$$
where $S$ and $\phi$ are momentum and buoyancy `\textit{shape factors}', respectively, defined as
\begin{equation}
	S = 1 + \frac{2}{b^{2}W^{2}}\int_{0}^{b} r\left(\bar{w}-W\right)^{2}\dd r,
	\label{eqn:S defn}
\end{equation}
and
\begin{equation}
	\phi = 1+\frac{2}{b^{2}Wg'}\int_{0}^{b} r\left(\bar{w}-W\right)\left(\bar{g_{r}}-G'\right)\dd r.
	\label{eqn:phi defn}
\end{equation}
Thus, the shape factors quantify the effect of non-uniform radial profiles of the axial velocity and density on the rates of transport of momentum and scalar species.  Crucially, we note that $S\ge 1$, which is representative of the more rapid transport of momentum than mass in the plume, unless the mean axial velocity is radially invariant within the plume (in which case $S\equiv 1$ and $\phi\equiv 1$).
 
The plume edge, given by the surface $r=b(z,t)$, is chosen to be at a radial distance such that the boundary terms in \eqref{eqn:full int eqns} are negligible in comparison to the integral terms (e.g. $\bar{w}(b) \ll W$, $\overline{u'w'}(b) \ll W^{2}$, etc.).  Furthermore, the entrainment assumption of \citet{Morton1956} allows the entrainment velocity to be written as $u_{e} = -kW$ where $k$ is the entrainment coefficient.  This simplification of the turbulence mixing dynamics assumes a similarity of turbulent structures at each height in the plume.  The entrainment coefficient must be determined empirically and, for fully developed plumes far from the source, a constant value $k=0.1\pm 0.01$ is appropriate \citep{Morton1956,List1982,Woods2010}.  However, near to the source (or for non-ideal initial conditions), the entrainment coefficient may vary substantially as the flow develops \citep{Wang2002,Kaminski2005,Ezzamel2015}.  In this study we take a constant entrainment coefficient (with the exception of appendix \ref{sec:app energy closure} where we also examine the steady solutions of a model with an entrainment coefficient that varies as the flow develops).

We note from \eqref{eqn:S defn} that the value of the momentum shape factor is tied to the choice of the plume width, $b(z,t)$.  If we assume the mean axial velocity has a self-similar Gaussian profile in the radial direction, as inferred from laboratory experiments on steady plumes \citep{Papanicolaou1988,Shabbir1994}, so that the mean axial velocity of the plume can be written as
\begin{equation}
	\bar{w}\left(r,z,t\right) = \overline{W}(z,t)\e^{-r^{2}/R^{2}},
	\label{eqn:w Gaussian}
\end{equation}
where $\overline{W}(z,t)$ is the mean axial velocity on the plume axis and $R(z,t)$ is a characteristic length scale for radial variation, then the shape factor is given by
\begin{equation}
	S = \frac{b^{2}\left(1+\e^{-b^{2}/R^{2}}\right)}{2R^{2}\left(1-\e^{-b^{2}/R^{2}}\right)}.
	\label{eqn:S Gaussian b}
\end{equation}
The plume width can be related to the length scale of the Gaussian radial profile $R(z,t)$ through a threshold on the mean axial velocity.  For example, \citet{Morton1956} and \citet{Papanicolaou1988} define the characteristic length scale of a Gaussian plume as the radial position at which the axial velocity is a factor of $1/\e$ of the centreline value, and this results in $S=1.08$.

The entrainment hypothesis and use of shape factors for the integral fluxes of momentum and buoyancy allow the integral conservation equations \eqref{eqn:full int eqns} to be written in terms of the integral fluxes of volume, momentum and buoyancy as,
\begin{subeqnarray}
	& & \fpd{}{t}\left(\frac{Q^{2}}{M}\right) + \fpd{Q}{z} = 2k\sqrt{M}, \slabel{eqn:int mass} \\[3pt]
	& & \fpd{Q}{t} + \fpd{}{z}\left(SM\right) = \frac{QF}{M} - \fpd{}{z}\int_{0}^{b} 2r\overline{{w'}^{2}}\dd r, \slabel{eqn:int mom} \\[3pt]
	& & \fpd{}{t}\left(\frac{QF}{M}\right) + \fpd{}{z}\left(\phi F\right) = \frac{b^{2}g}{\rho_{0}}\left(\fpd{\rho_{a}}{t} + W\fpd{\rho_{a}}{z}\right) - \fpd{}{z}\int_{0}^{b} 2r\overline{g_{r}'w'}\dd r. \slabel{eqn:int buoy}
	\label{eqn:int eqns}
\end{subeqnarray}
While the system \eqref{eqn:int eqns} applies to a spatially and temporally varying ambient density field, for the remainder of this study we consider plumes in an unstratified ambient, with $\rho_{a}$ constant.  The shape factors $S$ and $\phi$ could be spatially and temporally varying, as the radial profiles of the axial plume velocity and reduced gravity evolve \citep{Carazzo2006,Craske2015a,Craske2015b,Ezzamel2015}.  However, for fully developed plumes, laboratory experiments suggest that the axial plume velocity and reduced gravity attain self-similar forms \citep{Morton1956,Papanicolaou1988,Shabbir1994,Ezzamel2015} such that the shape factors can be taken to be constants.   Here we consider the simplest case of constant $S$ and $\phi\equiv 1$.

In the integral equations for conservation of momentum \eqref{eqn:int mom} and conservation of buoyancy \eqref{eqn:int buoy} we have retained integral terms that represent turbulent axial diffusion of momentum and buoyancy, respectively.  Typically, in steady plume models these diffusive terms have been neglected (we refer to the system of integral equations \eqref{eqn:int eqns} without the turbulent diffusive terms as the `non-diffusive' system).  Indeed, \citet{Morton1971} argues that
\begin{equation}
	\fpd{}{z}\left(\overline{{w'}^{2}}\right) \ll \frac{1}{r}\fpd{}{r}\left(r\overline{u'w'}\right), \quad \text{and} \quad \fpd{}{z}\left(\overline{w' {g_{r}}'}\right) \ll \frac{1}{r}\fpd{}{r}\left(r\overline{u' {g_{r}}'}\right),
\end{equation}
and therefore the contributions of the turbulent diffusion terms to the plume dynamics are of a similar magnitude to boundary terms that are neglected.  We note some studies on momentum driven jets from maintained sources retain the axial derivatives of quadratic fluctuation terms \citep[see e.g.][]{Shabbir1994,Wang2002,Yannopoulos2006}.  When these turbulent diffusion terms are neglected, the steady integral equations with pure plume boundary conditions ($Q=0$, $M=0$, $F=F_{0}$ at $z=0$) have well-known power-law solutions \citep{Morton1956}, and these solutions are little altered by the momentum shape factor; for $S\ge 1$ the steady solutions of the non-diffusive system of equations are
\begin{equation}
	Q = Q_{0}(z) = q_{0}z^{5/3}, \qquad M = M_{0} = m_{0} z^{4/3}, \qquad F = F_{0},
	\label{eqn:steady soln}
\end{equation}
where
\begin{equation}
	q_{0} = \frac{6k}{5}\left(\frac{9 k}{10}\right)^{1/3}\left(\frac{F_{0}}{S}\right)^{1/3}, \qquad m_{0} = \left(\frac{9 k}{10}\right)^{2/3}\left(\frac{F_{0}}{S}\right)^{2/3}.
	\label{eqn:steady coeffs}
\end{equation}
In the limit $S\to 1$ the solution of \citet{Morton1956} is recovered.  Furthermore, the effective radius of the plume, $b_{0}(z) = Q_{0}/\sqrt{M_{0}} = 6k z/5$, is independent of the shape factor.  Therefore, the momentum shape factor cannot be determined from measurement of the plume radius alone, in contrast to the entrainment coefficient that can be determined using the spreading rate of the steady plume.

\citet{Scase2012} advocate modelling of the turbulent diffusion terms (particularly the diffusion of momentum) in \eqref{eqn:int eqns} in the unsteady plume model to obtain a well-posed system of equations.  Recently, \citet{Craske2015a,Craske2015b} have demonstrated the relatively weak role of turbulent diffusion for momentum driven jets, but that diffusive effects in the jet occur due to the departure of flow variables from self-similar forms \citep[type II dispersion in the nomenclature of][]{Craske2015a,Craske2015b}.  However, we suggest that the dominant dynamics for turbulent plumes can be described by the non-diffusive system of equations.  Indeed, we show below that the inclusion of diffusive terms modelling turbulent diffusion can lead to difficulties in the integral model.  We therefore analyse the non-diffusive system and show that a momentum shape factor that differs from unity is sufficient to obtain a well-posed model.


\section{Difficulties associated with the turbulent diffusive terms}
\label{sec:Scase problems}

In an analysis of a non-diffusive unsteady plume model with shape factors $S\equiv 1$ and $\phi\equiv 1$, corresponding to top-hat radial profiles for the axial velocity and reduced gravity (i.e. $\bar{w}$ and $\bar{g_{r}}$ are radially invariant for $r\le b$ and equal to zero for $r>b$), \citet{Scase2012} assess the well-posedness of the system of equations by introducing small harmonic perturbations to the source buoyancy flux and examine the growth of the perturbations downstream of the source.  A linear analysis shows that the perturbations grow with distance from the source, indicating instability, and, more importantly, the growth rate of the perturbations increases without bound as the frequency of the harmonic oscillation of the source buoyancy flux increases \citep{Scase2012}.  The non-diffusive system of equations with $S\equiv 1$ and $\phi\equiv 1$ are therefore ill-posed, as there is a loss of continuous dependence of the solution on the boundary conditions \citep{Joseph1990,Scase2012}.  The ill-posedness is manifest in numerical solutions of the system of equations by an inability to compute solutions that are independent of the truncation implicit in the numerical scheme (e.g. grid-scale dependence for finite-difference methods).

The ill-posedness identified by \citet{Scase2012} in the non-diffusive system when top-hat profiles are assumed (i.e. when $S\equiv 1$) may be due to a missing physical process that provides a mechanism to curtail the unbounded growth of the arbitrarily short-wavelength/high frequency modes \citep{Joseph1990,Scase2012}.  In an attempt to regularize the ill-posed system, \citet{Scase2012} introduce a phenomenological model for the diffusive term in the momentum balance \eqref{eqn:int mom}, appealing to Prandtl's mixing length theory for turbulent eddy diffusion,
\begin{equation}
	\fpd{}{z}\int_{0}^{b} 2r\overline{{w'}^{2}}\dd r \approx -\frac{\kappa}{2k}b^{2}\fpd{}{z}\left[b W\fpd{W}{z}\right],
	\label{eqn:Scase diff term}
\end{equation}
where $\kappa>0$ is a dimensionless parameter characterizing the diffusion of momentum through the action of turbulent eddies whose length scale is set by the radius of the plume.  The equation expressing the balance of axial momentum proposed by \citet{Scase2012} is then given by
\begin{equation}
	\fpd{Q}{t}\ + S\fpd{M}{z} = \frac{QF}{M} + \frac{\kappa}{2k}\frac{Q^{2}}{M}\fpd{}{z}\left[\sqrt{M}\fpd{}{z}\left(\frac{M}{Q}\right)\right].
	\label{eqn:reg mom}
\end{equation}
\citet{Scase2012} provide numerical evidence that the diffusive term leads to a well-posed system of equations.  Note that, while \citet{Scase2012} take $S\equiv 1$, in the analysis presented below we analyse the more general problem of $S\ge 1$.

\begin{figure}
	\SetLabels
	\L (-0.07*0.43) \begin{sideways} $\Real{\alpha}$ \end{sideways} \\
	\L (0.275*0.90) (a) \\
	\L (0.61*0.90) (b) \\
	\L (0.955*0.90) (c) \\
	\R (0.0*0.95) $5$ \\
	\R (0.0*0.86) $4$ \\
	\R (0.0*0.77) $3$ \\
	\R (0.0*0.682) $2$ \\
	\R (0.0*0.595) $1$ \\
	\R (0.0*0.508) $0$ \\
	\R (0.0*0.416) $-1$ \\
	\R (0.0*0.328) $-2$ \\
	\R (0.0*0.236) $-3$ \\
	\R (0.0*0.151) $-4$ \\
	\R (0.0*0.065) $-5$ \\
	\L (0.15*-0.05) $\kappa$ \\
	\L (0.002*0.01) $0$ \\
	\L (0.08*0.01) $5$ \\
	\L (0.15*0.01) $10$ \\
	\L (0.225*0.01) $15$ \\
	\L (0.30*0.01) $20$ \\
	\L (0.5*-0.05) $\kappa$ \\
	\L (0.34*0.01) $0$ \\
	\L (0.417*0.01) $5$ \\
	\L (0.485*0.01) $10$ \\
	\L (0.562*0.01) $15$ \\
	\L (0.64*0.01) $20$ \\
	\L (0.85*-0.05) $\kappa_{1}$ \\
	\L (0.68*0.01) $0$ \\
	\L (0.757*0.01) $5$ \\
	\L (0.825*0.01) $10$ \\
	\L (0.902*0.01) $15$ \\
	\L (0.98*0.01) $20$ \\
	\endSetLabels
	\centerline{\strut\AffixLabels{\includegraphics[width=0.9\textwidth,keepaspectratio]{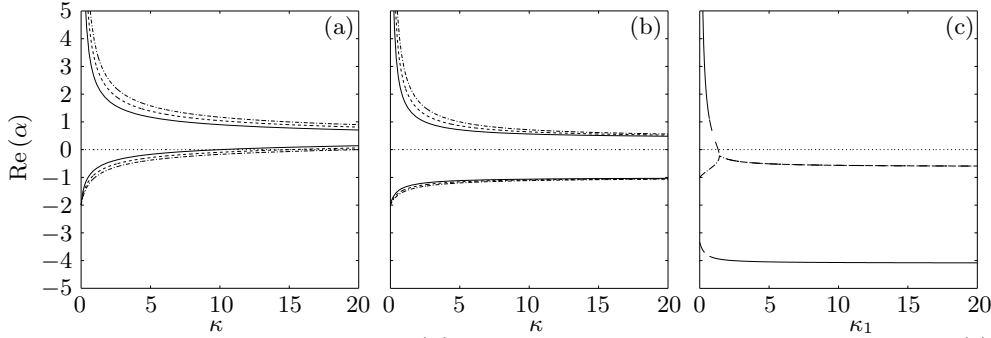}}}
	\caption{Spatial growth rates $\Real{\alpha}$ as a function of the diffusion coefficient for (a) the \citet{Scase2012} system with momentum diffusion \eqref{eqn:reg mom}, (b) the alternative form of momentum diffusion \eqref{eqn:reg mom 2}, and (c) diffusion of buoyancy \eqref{eqn:buoy diff}.  In (a) and (b) different values of the momentum shape factor are shown, with $S=1$ (solid line), $S=1.5$ (dashed line) and $S=2$ (dash-dot line).  In (c) the three branches of the growth rates are distinguished, with one branch corresponding to a real mode (solid line), and two branches that form a complex conjugate pair (dashed and dash-dot lines).}
	\label{fig:SpatialStabilityPlots}
\end{figure}

\citet{Scase2012} show that steady power-law solutions for pure plume boundary conditions exist for the diffusive system (with $\kappa>0$) and the structural change to the system of equations leads to a small modification of the solutions of \citet{Morton1956} (given by \ref{eqn:steady soln} with $S=1$).  The steady solutions of the `regularized' system of equations with turbulent diffusion of momentum are given by \citep{Scase2012}
\begin{subeqnarray}
	Q^{(\mathrm{SH})}_{0} &=& q_{0}\left(1-\frac{\kappa}{10S}\right)^{-1/3} z^{5/3}, \\[3pt]
	M^{(\mathrm{SH})}_{0} &=& m_{0}\left(1-\frac{\kappa}{10S}\right)^{-2/3} z^{4/3}, \\[3pt]
	F^{(\mathrm{SH})}_{0} &=& F_{0}.
	\label{eqn:Scase reg steady solns}
\end{subeqnarray}
However, we find that these power-law solutions are spatially unstable.  Indeed, taking perturbations of the form
\refstepcounter{equation}
$$
	Q^{(\mathrm{SH})}(z) = Q^{(\mathrm{SH})}_{0}(z)\left(1+\epsilon Q^{(\mathrm{SH})}_{1}(z)\right), \quad M^{(\mathrm{SH})}(z) = M^{(\mathrm{SH})}_{0}(z)\left(1+\epsilon M^{(\mathrm{SH})}_{1}(z)\right),
	\eqno{(\theequation{\mathit{a},\mathit{b}})}
	\label{eqn:spatial perturb}
$$
(noting that the buoyancy flux $F(z)\equiv F_{0}$ for a plume in an unstratified ambient) where $\epsilon>0$ is an ordering parameter, and linearizing the governing ordinary differential equations (for $\epsilon\ll 1$) we obtain,
\begin{subeqnarray}
	& & z\fd{Q^{(\mathrm{SH})}_{1}}{z} + \frac{5}{3}Q^{(\mathrm{SH})}_{1} - \frac{5}{6}M^{(\mathrm{SH})}_{1} = 0, \\[3pt]
	& & \frac{6\kappa}{5S}z^{2}\left(\fdd{M^{(\mathrm{SH})}_{1}}{z} - \fdd{Q^{(\mathrm{SH})}_{1}}{z}\right) - 2\left(1+\frac{\kappa}{10S}\right)z\fd{M^{(\mathrm{SH})}_{1}}{z} \nonumber \\
	& & \qquad + \frac{8}{3}Q^{(\mathrm{SH})}_{1} + 4\left(\frac{\kappa}{10S}-\frac{4}{3}\right)M^{(\mathrm{SH})}_{1} = 0.
\end{subeqnarray}
Seeking solutions of the form $Q^{(\mathrm{SH})}_{1} = q_{1}z^{\alpha}$ and $M^{(\mathrm{SH})}_{1} = m_{1}z^{\alpha}$, we find non-trivial solutions are possible if $\alpha=\alpha_{1}=-1$ or
\begin{equation}
	\alpha = \alpha_{\pm} = \frac{5S+4\kappa\pm\sqrt{25S^{2}+240S\kappa-4\kappa^{2}}}{6\kappa}.
	\label{eqn:alpha scase}
\end{equation}
At least one of $\alpha_{\pm}>0$ for $\kappa>0$ for any $S\ge 1$ (figure \ref{fig:SpatialStabilityPlots}a) and therefore the steady solutions of the \citet{Scase2012} 'regularized' model are spatially unstable.  We note that, in the limit $\kappa\to 0$, we find $\alpha_{+}$ no longer appears in the analysis while $\alpha_{-}\to -10/3$ and therefore the steady solutions become spatially stable (as expected, since the `regularized' system reduces to the classical \citet{Morton1956} plume model).  Spatial instability of the steady solutions is also found in a stratified ambient (not shown here) and in this case leads to severe difficulties as steady solutions cannot, in general, be found analytically and the spatial instability precludes the numerical computation of steady solutions.
Therefore, while the diffusive term introduced by \citet{Scase2012} into the momentum balance resolves the ill-posedness in the unsteady plume model, the modification of the governing equations renders the steady solutions spatially unstable.  While our analysis here considers pure plume boundary conditions (i.e. $Q(0)=0$, $M(0)=0$ and $F(0)=F_{0}$) in appendix \ref{sec:app phase plane} we show that spatially stable steady solutions with arbitrary initial conditions are not possible.

An alternative (although similar) form of the diffusion term can be formed by re-ordering the cross-sectional integration following the mixing-length parameterisation of the fluctuation vertical momentum flux (with boundary terms introduced that are assumed small and subsequently neglected), so that we write
\begin{equation}
	\fpd{}{z}\int_{0}^{b} 2r\overline{{w'}^{2}}\dd r \approx -\frac{\kappa}{2k}\fpd{}{z}\left[b^{3} W\fpd{W}{z}\right],
	\label{eqn:Alt diff term}
\end{equation}
in place of \eqref{eqn:Scase diff term}.  We then obtain the following expression for conservation of momentum,
\begin{equation}
	\fpd{Q}{t} + S\fpd{M}{z} = \frac{QF}{M} + \frac{\kappa}{2k}\fpd{}{z}\left[\frac{Q^{2}}{\sqrt{M}}\fpd{}{z}\left(\frac{M}{Q}\right)\right].
	\label{eqn:reg mom 2}
\end{equation}
The corresponding steady solutions are,
\begin{subeqnarray}
	Q^{(\mathrm{SH2})}_{0} &=& q_{0}\left(1+\frac{\kappa}{5S}\right)^{-1/3} z^{5/3}, \\[3pt]
	M^{(\mathrm{SH2})}_{0} &=& m_{0}\left(1+\frac{\kappa}{5S}\right)^{-2/3} z^{4/3}, \\[3pt]
	F^{(\mathrm{SH2})}_{0} &=& F_{0},
	\label{eqn:Scase reg 2 steady solns}
\end{subeqnarray}
and a spatial stability analysis gives growth rates $\alpha=\alpha_{1}=-1$ or
\begin{equation}
	\alpha = \alpha_{\pm} = \frac{5S-3\kappa\pm\sqrt{25S^{2}+170S\kappa+49\kappa^{2}}}{10\kappa}.
	\label{eqn:kappa scase 2}
\end{equation}
Thus, as with the \citet{Scase2012} momentum diffusion term, there is a growing mode (figure \ref{fig:SpatialStabilityPlots}b) and the steady solutions are spatially unstable.

Finally, we consider the change to the classical steady solutions of \citet{Morton1956} that occurs when a diffusive term is included in the equation for conservation of buoyancy \eqref{eqn:int buoy} rather than in the equation for conservation of momentum (i.e. we now take $\kappa=0$).  We again appeal to the Prandtl mixing length theory to model the fluctuation of the flux of buoyancy and obtain the following diffusive equation for the conservation of buoyancy,
\begin{equation}
	\fpd{}{t}\left(\frac{QF}{M}\right) + \fpd{F}{z} = \frac{\kappa_{1}}{2k}\fpd{}{z}\left[\frac{Q^{2}}{\sqrt{M}}\fpd{}{z}\left(\frac{F}{Q}\right)\right],
	\label{eqn:buoy diff}
\end{equation}
where $\kappa_{1}$ is a dimensionless parameter that describes the strength of the diffusion of buoyancy.  The classical steady solutions \eqref{eqn:steady soln} for pure plume boundary conditions remain as solutions of the system (and other possible solutions cannot satisfy the pure plume boundary condition at $z=0$).  However, a spatial stability analysis of the steady solutions (here including perturbations to the buoyancy flux in addition to the mass and momentum fluxes) shows that small amplitude perturbations to the steady solutions are amplified unless the magnitude of the turbulence induced is relatively large ($\kappa_{1}>4/3$, figure \ref{fig:SpatialStabilityPlots}c).

The consequences of these analyses are that, while axial diffusivity affects may capture some important unsteady processes in the mixing of momentum or buoyancy, the form of the parameterization significantly alters the steady states attainable by the system.  Indeed, the states equivalent to those established by \citet{Morton1956} have become spatially unstable and therefore the unsteady model with axial diffusivity is not able to describe the steady states.


\section{Well-posedness of the hyperbolic unsteady plume model}
\label{sec:well posed}

As the turbulent diffusive terms modelled phenomenologically using a Prandtl mixing-length approach leads to a pathology in the integral plume model whereby the well-established steady states of \citet{Morton1956} cannot be obtained, the regularisation through eddy diffusion, while potentially physically appealing, is mathematically problematic.  We therefore seek an alternative regularization of the system of unsteady integral equations; specifically, we examine the non-diffusive system of equations (i.e. we neglect the turbulent diffusive terms in \ref{eqn:int eqns}, as suggested by \citealp{Morton1971}) but allow the momentum shape factor to differ from unity to describe the different rates of transport of mass and momentum that results from non-uniform radial profiles.  Thus, from here on in we study the non-diffusive system of equations,
\refstepcounter{equation}
$$
	\fpd{}{t}\left(\frac{Q^{2}}{M}\right) + \fpd{Q}{z} = 2k\sqrt{M}, \quad \fpd{Q}{t} + S\fpd{M}{z} = \frac{QF}{M},  \quad \fpd{}{t}\left(\frac{QF}{M}\right) + \fpd{F}{z} = 0.
	\eqno{(\theequation{\mathit{a},\mathit{b},\mathit{c}})}
	\label{eqn:flux eqns}
$$

The system \eqref{eqn:flux eqns} is strictly hyperbolic when $S>1$, with three real characteristics wave speeds given by
\begin{equation}
	c_{0} = M/Q, \quad \text{and} \quad c_{\pm} = \left(S\pm \sqrt{S\left(S-1\right)}\right)M/Q,
\end{equation}
We note, however, in the limit $S\to 1$ there is a loss of hyperbolicity since the eigenvectors of the characteristic equation for the system \eqref{eqn:flux eqns} do not span $\mathbb{R}^{3}$, although the wave speeds remain real and equal to $c_{0}$; the system is formally parabolic for $S=1$ \citep{Scase2006a}.  As we demonstrate below, the change in the characteristic structure, and so change in type, of the governing equations that occurs for $S=1$ fundamentally alters solutions of the system of equations.  (Note, if a shape factor for the buoyancy flux that differs from unity is included in \ref{eqn:flux eqns}c, then $c_{0}=\phi M/Q$ while $c_{\pm}$ remain unchanged, and a loss of hyperbolicity occurs if $\phi = S\pm\sqrt{S\left(S-1\right)}$.)

In order to establish well-posedness of the system of equations, we analyse the evolution of small perturbations to the steady solutions.  It is convenient for subsequent analysis and numerical computations to factor-out the steady solution and to introduce a new spatial coordinate, $x = \left(q_{0}/m_{0}\right)z^{4/3}$.  We consider the stability of the steady solution to small perturbations.  Therefore, we introduce an ordering parameter $\epsilon\ll 1$ and perturbations to the steady solution of the form,
\begin{subeqnarray}
	Q(x,t) &=& Q_{0}(x)\left(1+\epsilon q(x,t)\right), \\
	M(x,t) &=& M_{0}(x)\left(1+\epsilon m(x,t)\right), \\
	F(x,t) &=& F_{0}\left(1+\epsilon f(x,t)\right),
\end{subeqnarray}
and linearize the governing equations \eqref{eqn:flux eqns} to obtain
\begin{equation}
	\tensor{A}\fpd{\bsym{q}}{t} + \frac{4}{3}\tensor{B}\fpd{\bsym{q}}{x} + \tensor{C}\frac{1}{x}\bsym{q} = \bsym{0},
	\label{eqn:linearized system}
\end{equation}
where $\bsym{q} = \left(q,m,f\right)^{\mathrm{T}}$ and the matrices $\tensor{A}$, $\tensor{B}$, and $\tensor{C}$ are given by
\begin{equation}
	\tensor{A}= \begin{pmatrix} 2 & -1 & 0 \\ 1 & 0 & 0 \\ 1 & -1 & 1 \end{pmatrix}, \quad \tensor{B}= \begin{pmatrix} 1 & 0 & 0 \\  0 & S & 0 \\ 0 & 0 & 1 \end{pmatrix}, \quad \text{and} \quad
	\tensor{C}= \begin{pmatrix} 5/3 & -5/6 & 0 \\ -4S/3 & 8S/3 & -4S/3 \\ 0 & 0 & 0 \end{pmatrix}.
	\label{eqn:lin matrices}
\end{equation}

When $S\equiv 1$ the system \eqref{eqn:flux eqns} was shown to be ill-posed by \citet{Scase2012} who examined the response of the linearized system of equations to small amplitude harmonic variation of frequency $\omega$ in the source buoyancy flux, and found a closed form solution for the perturbations in terms of special functions.  It was shown that the amplitude of the perturbations grows with distance from the source as $\omega^{-7/8}x^{-7/8}\exp\left\{\sqrt{15\omega x}/4\right\}$ for $x \gg 1$ \citep{Scase2012}.  Thus, the steady solutions are linearly unstable, as the amplitude of the perturbations grows as they propagate away from the source and, furthermore, the system of equations \eqref{eqn:flux eqns} with $S=1$ is ill-posed since high frequency fluctuations increase in amplitude most rapidly with no cut-off \citep{Scase2012}.

For $S>1$ no closed form solution of the linear system \eqref{eqn:linearized system} can be obtained.  We therefore examine the evolution of perturbations in the far-field through an asymptotic analysis of the linear system \eqref{eqn:linearized system} in the neighbourhood of the irregular singular point $x\to\infty$ \citep{BenderOrszag}.  We consider two stability problems; an initial value problem where the evolution of an arbitrary initial perturbation with compact support,
\begin{equation}
	\bsym{q}(x,0) = \bsym{q}_{0}(x) \quad \text{with} \quad \bsym{q}_{0}(x)\to 0 \text{ as } x\to\infty,
\end{equation}
is investigated, and a boundary value problem where a fluctuation at the source $x=0$ is imposed and the response of the system downstream of the source is examined.  Both of these problems can be analysed conveniently through the use of integral transforms of the linear system \eqref{eqn:linearized system}; a Laplace transform in time for the initial value problem and a Fourier transform in time for the boundary value problem.  Denoting the Laplace transform of $\bsym{q}(x,t)$ as $\skew3\hat{\bsym{q}}(x,p)$ (with the Fourier transform obtained by taking $p=\ci\omega$ where $\omega$ is the frequency of the harmonic source fluctuation imposed in the boundary value problem) we obtain,
\begin{equation}
	p\tensor{A}\skew3\hat{\bsym{q}} + \frac{4}{3}\tensor{B}\fd{\skew3\hat{\bsym{q}}}{x} + \frac{1}{x}\tensor{C}\skew3\hat{\bsym{q}} = \tensor{A}\bsym{q}_{0},
	\label{eqn:LT linear sys}
\end{equation}
The far-field behaviour is obtained conveniently by letting $\skew3\hat{\bsym{q}} = \bsym{f}(x)\e^{g(x)}$ \citep{BenderOrszag} with $g(x)$ a singular function and $\bsym{f}(x)$ regular as $x\to \infty$, so $\bsym{f}(x) = \bsym{f}_{0} + \bsym{f}_{1}x^{-1} + \bsym{f}_{2}x^{-2} + \dots$ for $x\gg 1$.  The linear system \eqref{eqn:LT linear sys} can then be written as
\begin{equation}
	\frac{4}{3}\tensor{B}\fd{\bsym{f}}{x} + \left(p\tensor{A} + \frac{4}{3}\tensor{B}\fd{g}{x} + \frac{1}{x}\tensor{C}\right)\bsym{f} = 0,
	\label{eqn:far-field eqn}
\end{equation}
A leading-order balance requires $g(x) \sim p\lambda x$ as $x \to \infty$.  Then $\lambda$ and $\bsym{f}_{0}$ satisfy the generalized eigenproblem $\tensor{A}\bsym{f}_{0} = -4\lambda\tensor{B}\bsym{f}_{0}/3$ and therefore,
\begin{subeqnarray}
	\lambda &=& \lambda_{0} = -\frac{3}{4}, \quad \text{with} \quad \bsym{f}_{0} = \bsym{f}_{00} = \left(0,0,1\right)^{\mathrm{T}}, \quad \bsym{f}^{L}_{0} = \bsym{f}^{L}_{00} = \left(-1,0,1\right), \\[3pt]
	\lambda &=& \lambda_{+} = -\frac{3}{4}\left(\frac{S+\sqrt{S\left(S-1\right)}}{S}\right), \quad \text{with} \quad \bsym{f}_{0} = \bsym{f}_{0+} = \left(1,\frac{1-\left(S-\sqrt{S\left(S-1\right)}\right)}{\sqrt{S\left(S-1\right)}},1\right)^{\mathrm{T}}, \nonumber \\[3pt]
	& & \qquad \qquad \bsym{f}^{L}_{0} = \bsym{f}^{L}_{0+} = \left(S+\sqrt{S\left(S-1\right)},-1,0\right), \\[3pt]
	\lambda &=& \lambda_{-} = \frac{3}{4}\left(\frac{S-\sqrt{S\left(S-1\right)}}{S}\right), \quad \text{with} \quad \bsym{f}_{0} = \bsym{f}_{0-} = \left(1,-\frac{1-\left(S+\sqrt{S\left(S-1\right)}\right)}{\sqrt{S\left(S-1\right)}},1\right)^{\mathrm{T}}, \nonumber \\[3pt]
	& & \qquad \qquad \bsym{f}^{L}_{0} = \bsym{f}^{L}_{0-} = \left(S-\sqrt{S\left(S-1\right)},-1,0\right),
	\label{eqn:leading order}
\end{subeqnarray}
where $\bsym{f}^{L}_{0}$ denotes the left eigenvector satisfying $\bsym{f}^{L}_{0}\tensor{A} = -4\lambda\bsym{f}^{L}_{0}\tensor{B}/3$.

To proceed further we let $g(x) \sim p\lambda x + \mu \log x$ for $x\gg 1$.  Substitution into \eqref{eqn:far-field eqn} and balancing coefficients of $x$ gives, at order $\textit{O}\left(1/x\right)$,
\begin{equation}
	p\left(\tensor{A}+\frac{4}{3}\lambda\tensor{B}\right)\bsym{f}_{1} +\left(\frac{4}{3}\mu\tensor{B} + \tensor{C}\right)\bsym{f}_{0} = \bsym{0},
\end{equation}
and so, by multiplying on the left by $\bsym{f}^{L}_{0}$, we find
\begin{equation}
	\mu = -\frac{3}{4}\frac{\bsym{f}_{0}^{L}\tensor{C}\bsym{f}_{0}}{\bsym{f}_{0}^{L}\tensor{B}\bsym{f}_{0}},
\end{equation}
and therefore
\begin{subeqnarray}
	\mu &=& \mu_{0} = 0 \mbox{ when $\lambda=\lambda_{0}$}, \slabel{eqn:mu0} \\[3pt]
	\mu &=& \mu_{+} = -\frac{13}{8} - \frac{5\left(2S-1\right)}{16\sqrt{S\left(S-1\right)}} \mbox{ when $\lambda=\lambda_{+}$}, \slabel{eqn:mu+} \\[3pt]
	\mu &=& \mu_{-} = -\frac{13}{8} + \frac{5\left(2S-1\right)}{16\sqrt{S\left(S-1\right)}} \mbox{ when $\lambda=\lambda_{+}$} \slabel{eqn:mu-}.
	\label{eqn:perts decay rates}
\end{subeqnarray}
The coefficient vectors ($\bsym{f}_{1}$, $\bsym{f}_{2}$, etc.) can be determined from higher order terms in the expansion but are not given here.

\begin{figure}
	\SetLabels
	\L (-0.09*0.5) $\mu_{\pm}$ \\
	\R (0.015*0.94) $5$ \\
	\R (0.015*0.855) $4$ \\
	\R (0.015*0.77) $3$ \\
	\R (0.015*0.68) $2$ \\
	\R (0.015*0.595) $1$ \\
	\R (0.015*0.505) $0$ \\
	\R (0.015*0.42) $-1$ \\
	\R (0.015*0.335) $-2$ \\
	\R (0.015*0.25) $-3$ \\
	\R (0.015*0.16) $-4$ \\
	\R (0.015*0.08) $-5$ \\
	\L (0.5*-0.03) $S$ \\
	\L (0.01*0.03) $1.0$ \\
	\L (0.10*0.03) $1.1$ \\
	\L (0.19*0.03) $1.2$ \\
	\L (0.285*0.03) $1.3$ \\
	\L (0.38*0.03) $1.4$ \\
	\L (0.47*0.03) $1.5$ \\
	\L (0.565*0.03) $1.6$ \\
	\L (0.66*0.03) $1.7$ \\
	\L (0.75*0.03) $1.8$ \\
	\L (0.845*0.03) $1.9$ \\
	\L (0.935*0.03) $2.0$ \\
	\endSetLabels
	\centerline{\strut\AffixLabels{\includegraphics[width=0.6\textwidth,keepaspectratio]{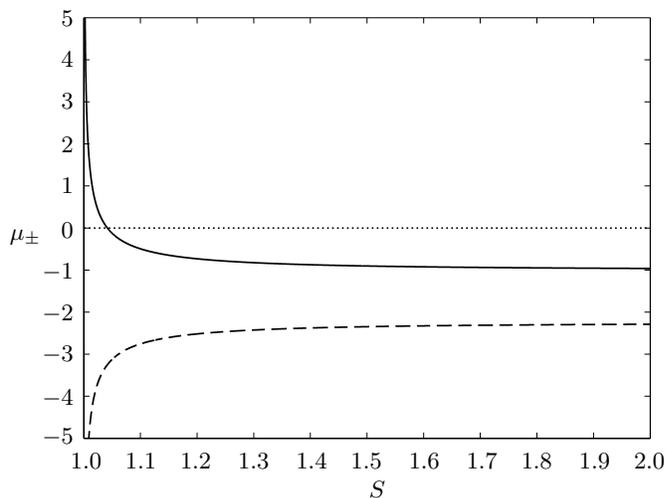}}}
	\caption{The exponents of the far-field algebraic growth rates of the amplitude of the volume flux perturbation, $\mu_{+}$ (dashed line) and $\mu_{-}$ (solid line), as functions of the shape factor $S$.  The perturbation to the volume flux grows with increasing distance from the source when $\mu_{-}>0$ which occurs for $S<25/24$.}
	\label{fig:mu plot}
\end{figure}

The leading order behaviour in the far-field is therefore given by,
\begin{equation}
	\bsym{q}(x,t) \sim \bsym{f}_{0}x^{\mu}\delta\left(t-\lambda x\right),
\end{equation}
corresponding to the propagation of discontinuities whose strength varies algebraically with distance from the source.  The amplitude of the perturbations grows if $1<S<25/24$ whereas the amplitude decays algebraically if $S>25/24$ (figure \ref{fig:mu plot}).  Importantly, the algebraic growth rates $\mu$ do not depend on the transform variable $p$.  Therefore, for the initial value problem the evolution of perturbations in the far-field is independent of the spatial length scale of the initial perturbation, whereas when considering a boundary value problem the growth rate of the perturbations do not depend on the frequency of the harmonic boundary oscillations.  Therefore, the far-field asymptotic analysis shows that the system \eqref{eqn:flux eqns} with $S>1$ is well-posed as there is a continuous dependence of the solutions on the initial or boundary data and small scales (either spatial scales in the case of an initial value problem or temporal scales for a boundary value problem) are not amplified more rapidly than longer scales.  This is in contrast to the far-field limit of the solution of a boundary value problem that is determined by \citet{Scase2012}, where exponential growth of the amplitude of perturbations are found with a growth rate that increases exponentially with $\sqrt{\omega}$, and therefore when $S=1$ the system of equations is ill-posed.  Numerical solutions demonstrating the evolution of a localised initial perturbation are presented in \S\ref{sec:numerical solutions} below.

\section{Solutions of the well-posed unsteady plume model}
\label{sec:numerical solutions}

We consider now solutions of the nonlinear system of equations \eqref{eqn:flux eqns} with $S>1$.  As the system of equations \eqref{eqn:flux eqns} is hyperbolic in this parameter regime, solutions may exhibit discontinuities, across which we enforce the following jump conditions that respectively impose the conservation of mass, momentum and buoyancy fluxes over a discontinuity at $z=z_s(t)$, moving with velocity $c=\dd z_s/\dd t$,
\begin{equation}
	\left[Q-c\frac {Q^{2}}{M}\right]_{z=z_{s}^{-}}^{z=z_{s}^{+}}=0,\quad
	\left[\vphantom{\frac{1}{1}} SM-cQ\right]_{z=z_{s}^{-}}^{z=z_{s}^{+}}=0\quad \text{and} \quad
	\left[F-c\frac {QF}{M}\right]_{z=z_{s}^{-}}^{z=z_{s}^{+}}=0.
	\label{eqn:jump conditions}
\end{equation}
These jump conditions admit two different types of discontinuous solutions, which are more easily analysed in terms of the primitive variables $b$, $W$ and $G'$.  Further denoting the values of the dependent variables either side of the discontinuity with superscripts $^{+}$ and $^{-}$, corresponding to $z=z_{s}^{+}$ and $z=z_{s}^{-}$, respectively, we find that provided the velocity is discontinuous $(W^{+}\neq W^{-})$ then the reduced gravity $G'$ is continuous (i.e $G'^{+}=G'^{-}$).  Furthermore in this case, by eliminating $b^{+}$ and $b^{-}$, we find that
\begin{equation}
	\left(W^{+}-W^{-}\right)\left(c^2-cS(W^{+}+W^{-})+SW^{+}W^{-}\right)=0.
\end{equation}
When the shape factor is equal to unity, the only solution is $c=W^{+}=W^{-}$ and this contradicts the assumption that there is a discontinuity.  Thus for $S=1$ it is not possible to construct discontinuous solutions.  However, when $S>1$, we find
\begin{equation}
	c=\textfrac{1}{2} S(W^{+} +W^{-})-\textfrac{1}{2} \left( S^2(W^{+} + W^{-})^2-4S W^{+}W^{-}\right)^{1/2},
	\label{eqn:shock speed}
\end{equation}
where the negative root has been chosen so that $W^{-}>c$, a condition required for causality.  The other type of discontinuity occurs when $W^{+}=W^{-}=c$.  Then the radius of the plume is also continuous, $b^{+}=b^{-}$, but the reduced gravity may be discontinuous and potentially even unbounded.  We refer to this latter form as a `contact' discontinuity, which occurs due to a linearly degenerate field in the system of equations \citep{Lax}.

To calculate numerically solutions of the nonlinear hyperbolic system of equations, we employ the non-oscillatory central scheme of \citet{Kurganov2000} with a third-order total variation diminishing Runge--Kutta time stepping scheme \citep{Gottlieb1998} using an adaptive time step that ensures the CFL number remains fixed at $1/8$ to maintain numerical stability.  The high-resolution central scheme is a shock-capturing numerical method for conservation laws \citep{Kurganov2000} and has been used extensively to compute solutions of nonlinear hyperbolic systems.  For numerical convenience, we factor-out the steady solutions from the dependent variables and compute solutions in the transformed spatial coordinate $x$ as, although the system is not autonomous in this representation, the characteristic wave speeds remain bounded near to the source (i.e. as $x\to 0$) when the steady solution is realized, whereas the wave speeds of the system \eqref{eqn:flux eqns} diverge as $z\to 0$ and therefore small time-steps are required to maintain numerical stability.

We consider first the evolution of a perturbation to the steady solution \eqref{eqn:steady soln} to confirm the far-field asymptotic analysis in \S\ref{sec:well posed}.  We take as an initial condition a Gaussian perturbation (in the transformed coordinate $x$) to the steady momentum flux of the form
\begin{equation}
	M(x,0)/M_{0}(x) = 1 + 0.01\e^{-10\left(x-5\right)^{2}},
\end{equation}
while the volume flux and buoyancy flux are not perturbed from the steady values, and take the shape factor to be $S=1.1$.  We compute numerically the evolution of the perturbation using the \citet{Kurganov2000} central scheme for the nonlinear system rather than using the linearised equations since the linearisation introduces a subtle structural change in the governing equations; for the linearised system, all field are (trivially) linearly degenerate and so discontinuities are of the contact discontinuity type \citep{Lax} whereas two fields in the nonlinear system are genuinely nonlinear.  We find that the \citet{Kurganov2000} scheme resolves accurately the shocks and rarefactions associated with the genuinely nonlinear fields, but has less accuracy when contact discontinuities occur for linearly degenerate fields \citep[see e.g.][]{KurganovPetrova2000}.  In order to track the evolution of the perturbations to long times (and so large distance from the source), we implement a moving numerical domain, using the characteristic wave speeds of the linearized system to advect the lower and upper grid points.  The numerical domain is then advected downstream and the number of grid points increases as the spatial extent of the perturbation grows in order to maintain a specified numerical resolution.

\begin{figure}
	\SetLabels
	\L (0.08*0.50) $(Q/Q_{0}-1)\times 10^{3}$ \\
	\L (0.4*0.50) $(M/M_{0}-1)\times 10^{3}$ \\
	\L (0.73*0.50) $(F/F_{0}-1)\times 10^{3}$ \\
	\L (-0.06*0.75) $x$ \\
	\L (0.05*0.52) $-5$ \\
	\L (0.275*0.52) $5$ \\
	\L (0.32*0.52) $-10$ \\
	\L (0.633*0.52) $10$ \\
	\L (0.7*0.52) $-5$ \\
	\L (0.925*0.52) $5$ \\
	\R (0.01*0.533) $0$ \\
	\R (0.01*0.578) $10$ \\
	\R (0.01*0.623) $20$ \\
	\R (0.01*0.668) $30$ \\
	\R (0.01*0.712) $40$ \\
	\R (0.01*0.757) $50$ \\
	\R (0.01*0.802) $60$ \\
	\R (0.01*0.847) $70$ \\
	\R (0.01*0.891) $80$ \\
	\R (0.01*0.936) $90$ \\
	\R (0.01*0.980) $100$ \\
	\L (0.08*-0.0) $(b/b_{0}-1)\times 10^{3}$ \\
	\L (0.4*-0.0) $(W/W_{0}-1)\times 10^{3}$ \\
	\L (0.73*-0.0) $(G'/G'_{0}-1)\times 10^{3}$ \\
	\L (-0.06*0.25) $x$ \\
	\L (-0.0*0.02) $-5$ \\
	\L (0.315*0.02) $5$ \\
	\L (0.335*0.02) $-5$ \\
	\L (0.418*0.02) $0$ \\
	\L (0.492*0.02) $5$ \\
	\L (0.56*0.02) $10$ \\
	\L (0.633*0.02) $15$ \\
	\L (0.67*0.02) $-4$ \\
	\L (0.755*0.02) $-2$ \\
	\L (0.867*0.02) $0$ \\
	\L (0.966*0.02) $2$ \\
	\R (0.01*0.035) $0$ \\
	\R (0.01*0.080) $10$ \\
	\R (0.01*0.124) $20$ \\
	\R (0.01*0.169) $30$ \\
	\R (0.01*0.213) $40$ \\
	\R (0.01*0.258) $50$ \\
	\R (0.01*0.303) $60$ \\
	\R (0.01*0.348) $70$ \\
	\R (0.01*0.392) $80$ \\
	\R (0.01*0.437) $90$ \\
	\R (0.01*0.481) $100$ \\
	\endSetLabels
	\centerline{\strut\AffixLabels{\includegraphics[width=0.85\textwidth,keepaspectratio]{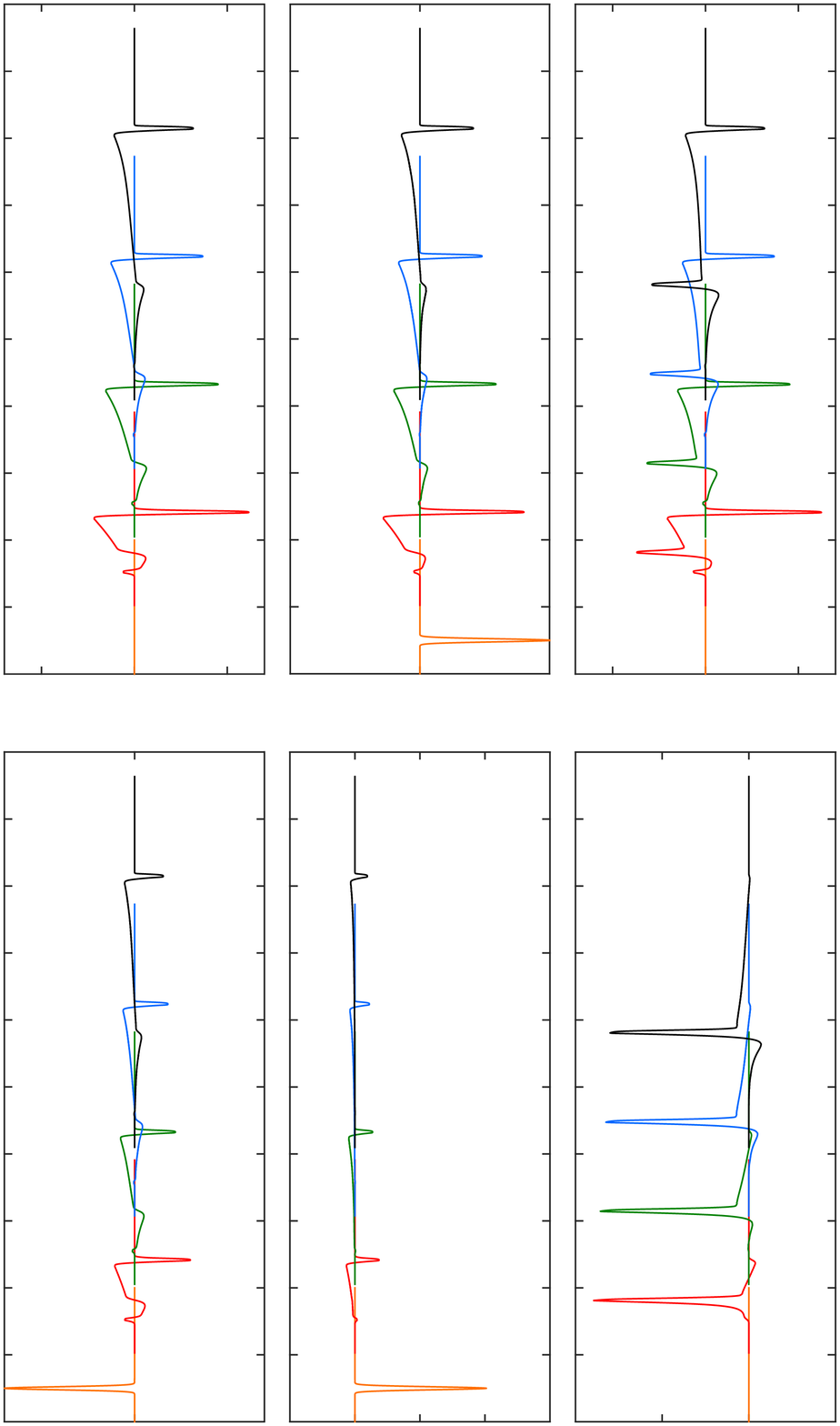}}}
	\caption{The perturbations to the fluxes of volume, $Q/Q_{0}-1$, momentum, $M/M_{0}-1$, and buoyancy, $F/F_{0}-1$, and to the plume radius, $b/b_{0}-1$, the plume velocity, $W/W_{0}-1$, and the reduced gravity, $G'/G'_{0}-1$, as functions of the scaled distance from the source, $x$, for dimensionless times $t=0$ (orange, colour online), $t=10$ (red, colour online), $t=20$ (green, colour online), $t=30$ (blue, colour online), and $t=40$ (black, colour online).  The momentum shape factor is fixed at a value $S=1.1$.  The numerical solution scheme adopts a moving and expanding spatial grid with a fixed grid spacing of $\Delta x = 0.01$.}
	\label{fig:StabProfiles_IVP}
\end{figure}

In figure \ref{fig:StabProfiles_IVP} we illustrate the evolution of the initial perturbation for dimensionless times $t\le 40$, showing the perturbation to the plume radius, mass flux and buoyancy flux as functions of the scaled spatial coordinate $x$.  The initial perturbation develops into a pulse whose spatial extent grows   as it propagates.  The amplitude of the perturbation slowly decreases, demonstrating linear stability of the system of equations for $S=1.1$.  The three discontinuities are apparent in the evolving perturbation, with the contact discontinuity that propagates at the speed of the intermediate characteristic most clearly seen in the perturbation to the reduced gravity, $G'$.

\begin{figure}
	\SetLabels
	\L (0.5*-0.0) $t$ \\
	\L (0.91*0.93) (a) \\
	\L (-0.05*0.75) $\mathcal{I}$ \\
	\L (0.17*0.53) $10$ \\
	\L (0.552*0.53) $100$ \\
	\L (0.93*0.53) $1000$ \\
	\L (0.0*0.61) $10^{-6}$ \\
	\L (0.0*0.79) $10^{-5}$ \\
	\L (0.0*0.97) $10^{-4}$ \\
	\L (0.91*0.45) (b) \\
	\L (-0.05*0.3) $\mathcal{I}$ \\
	\L (0.05*0.04) $0.1$ \\
	\L (0.355*0.04) $1$ \\
	\L (0.65*0.04) $10$ \\
	\L (0.94*0.04) $100$ \\
	\L (0.0*0.07) $10^{-6}$ \\
	\L (0.0*0.15) $10^{-5}$ \\
	\L (0.0*0.235) $10^{-4}$ \\
	\L (0.0*0.315) $10^{-3}$ \\
	\L (0.0*0.395) $10^{-2}$ \\
	\L (0.0*0.48) $10^{-1}$ \\
	\endSetLabels
	\centerline{\strut\AffixLabels{\includegraphics[width=0.85\textwidth,keepaspectratio]{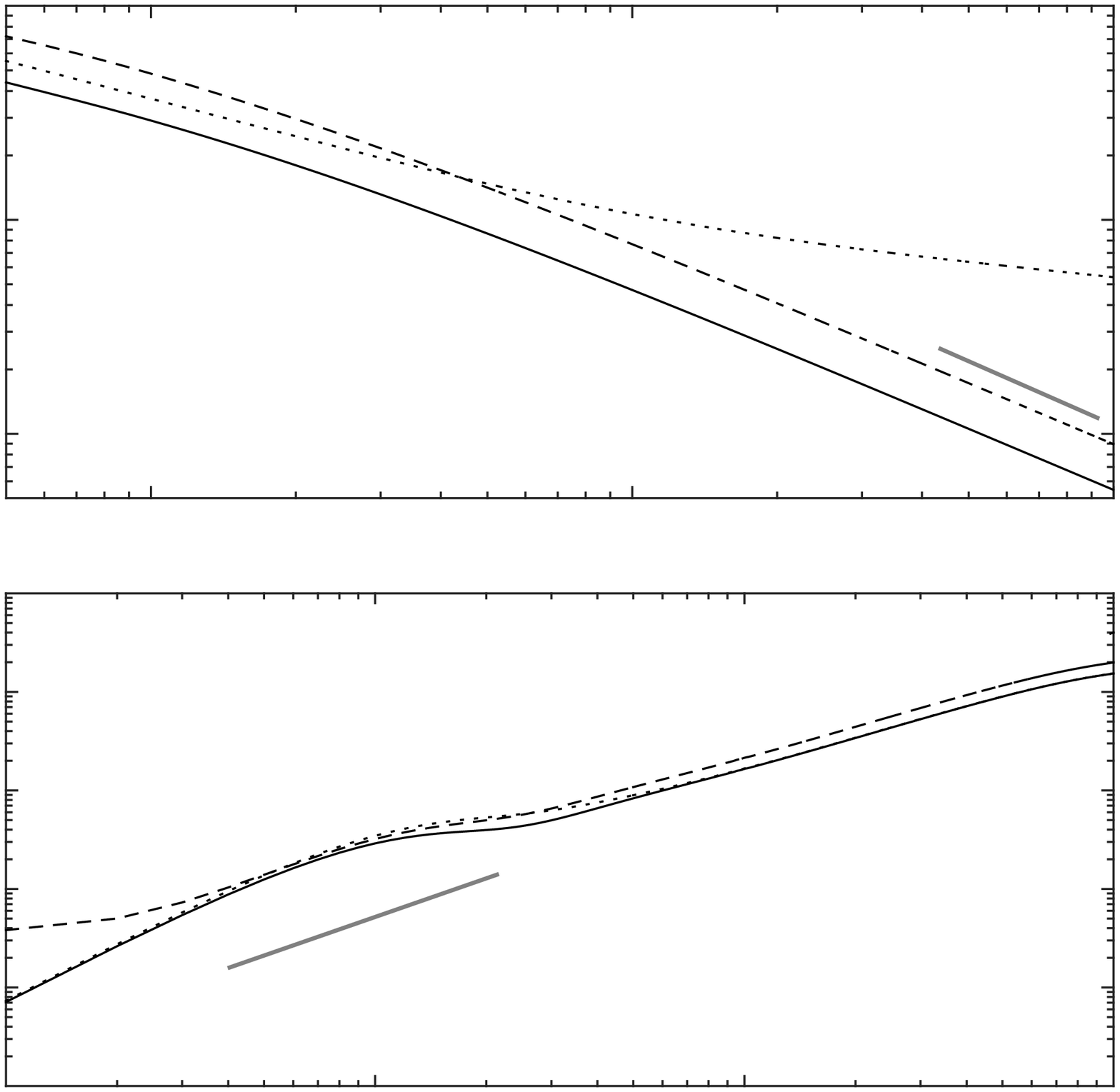}}}
	\caption{Integral measures of the size of perturbations from the steady solutions for the volume flux, $\mathcal{I}(Q)$ (solid line), momentum flux, $\mathcal{I}(M)$ (dashed line), and buoyancy flux, $\mathcal{I}(F)$ (dotted line) as functions of the dimensionless time $t$, for a shape factor (a) $S=1.1$ and (b) $S=1.02$.  The thick grey line illustrates the predicted growth rate of perturbations of $t^{2\mu_{+}}$ with (a) $\mu_{+}\approx -0.49$ for $S=1.1$ and (b) $\mu_{+}\approx 0.65$ for $S=1.02$ obtained from the far-field asymptotic analysis \eqref{eqn:mu+}.}
	\label{fig:StabInts}
\end{figure}

We introduce integral measures of the amplitude of the perturbations, with
\begin{equation}
	\mathcal{I}(Q) = \int_{0}^{\infty} \left(\frac{Q(x,t)}{Q_{0}(x)} - 1\right)^{2}\dd x,
\end{equation}
for the volume flux, and similarly for the fluxes of momentum and buoyancy.  Figure \ref{fig:StabInts}a shows the time evolution of $\mathcal{I}(Q)$, $\mathcal{I}(M)$ and $\mathcal{I}(F)$ when the shape factor $S=1.1$.  Following a short transient reorganization of the initial condition for $t<5$ (not shown on the figure) the perturbations to the steady solution decay as they propagate through the domain.
The perturbations to the steady fluxes of volume and momentum closely follow the prediction of the far-field analysis for $t>100$, at which time the signal of the perturbation that travels with the fastest characteristic has reached $x\approx 197$.  The decay in the perturbation to the buoyancy flux is less rapid and the rate of decay is diminishing as time progresses.  This is expected from the far-field asymptotic analysis of the linearised system where a component of the buoyancy flux is found to be advected at the speed of the plume without change in amplitude (see \ref{eqn:mu0}).

For $S<25/24$ our far-field asymptotic analysis shows that the steady solutions are linearly unstable.  Numerical solutions of the governing equations support the asymptotic analysis, as illustrated in figure \ref{fig:StabInts}b which shows the time evolution of the integral measures of the size of the perturbations for a shape factor $S=1.02$.  For early times ($t<10$) the perturbations to the steady solutions grow algebraically with a growth rate that is close to the growth rate predicted from the far-field analysis.  We note that at early times the perturbation has not propagated far downstream from the localised initial condition and therefore the far-field asymptotic analysis  would not be expected to describe fully the evolution.  At later times ($t>30$) the growth of the perturbations deviate substantially from the predictions of the linear far-field analysis as nonlinear effects begin to influence the evolution.

\begin{figure}
	\SetLabels
	\L (0.04*0.9) (a) \\
	\L (0.525*0.9) (b) \\
	\L (0.2*-0.02) $Q/Q_{0}-1$ \\
	\L (-0.07*0.5) $x$ \\
	\L (-0.01*0.04) $-0.03$ \\
	\L (0.25*0.04) $0$ \\
	\L (0.45*0.04) $0.03$ \\
	\L (0.7*-0.02) $Q/Q_{0}-1$ \\
	\L (0.55*0.04) $-0.04$ \\
	\L (0.74*0.04) $0$ \\
	\L (0.87*0.04) $0.04$ \\
	\R (0.01*0.08) $0$ \\
	\R (0.01*0.18) $5$ \\
	\R (0.01*0.27) $10$ \\
	\R (0.01*0.37) $15$ \\
	\R (0.01*0.47) $20$ \\
	\R (0.01*0.56) $25$ \\
	\R (0.01*0.655) $30$ \\
	\R (0.01*0.75) $35$ \\
	\R (0.01*0.85) $40$ \\
	\R (0.01*0.945) $45$ \\
	\endSetLabels
	\centerline{\strut\AffixLabels{\includegraphics[width=0.85\textwidth,keepaspectratio]{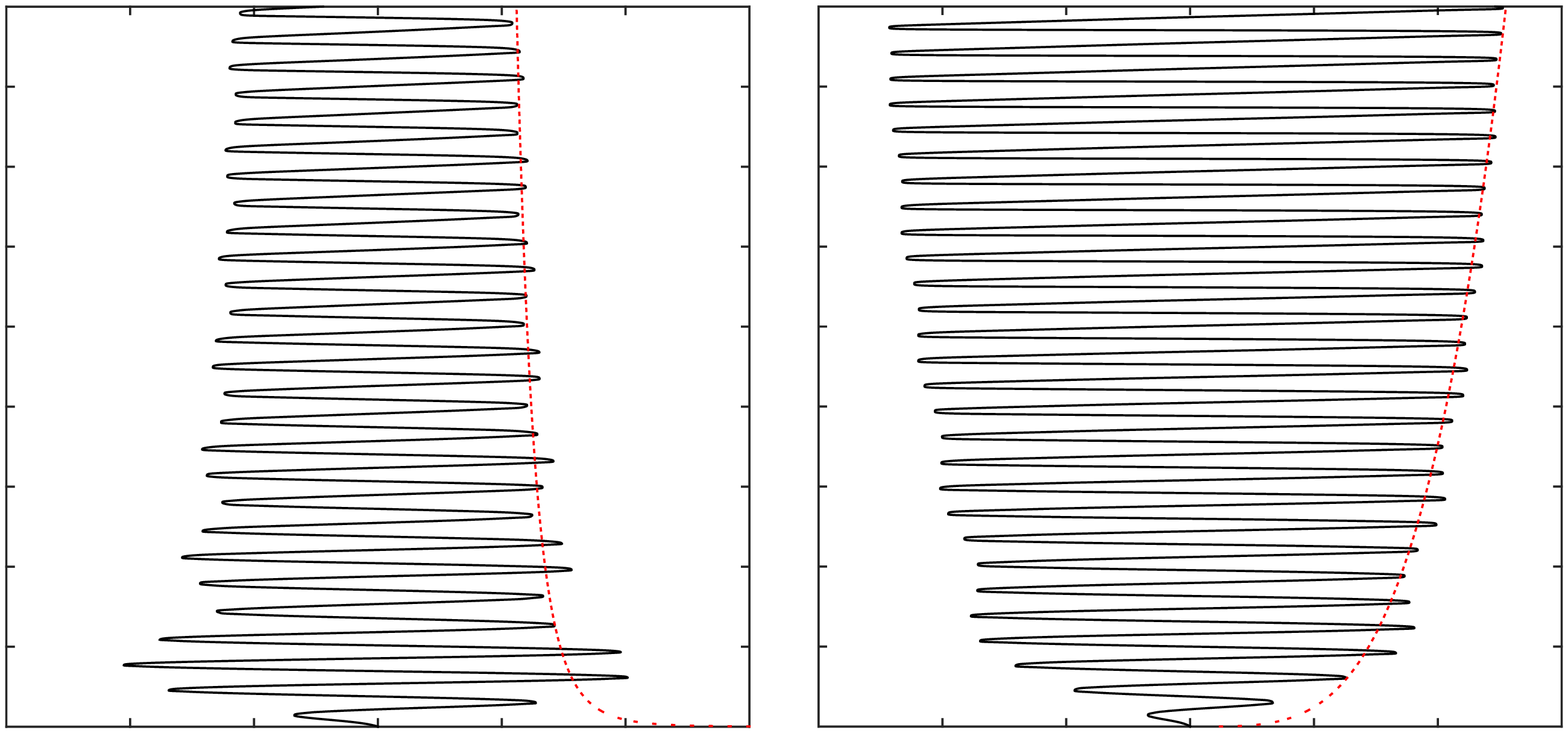}}}
	\caption{The perturbation to the steady volume flux due to harmonic oscillation to the source buoyancy flux with period $2\pi$ as a function of the scaled spatial coordinate $x$ at dimensionless time $t=99$ for a shape factor (a) $S=1.05$ and (b) $S=1.03$ .  The red dashed line (colour online) illustrates an algebraic growth of perturbations of the form $c x^{\mu}$ where (a) $\mu = -0.12$ for $S=1.05$ and (b) $\mu=0.26$ for $S=1.03$.  The prefactor $c$ is selected to provide an envelope of the numerical solution.}
	\label{fig:StabFluctuation}
\end{figure}

Numerical solutions of the governing equations also support the far-field asymptotic analysis of the linearized system when a harmonic oscillation of the source boundary conditions are imposed.  Figure \ref{fig:StabFluctuation} illustrates the spatial structure of the perturbation to the steady volume flux at time $t=99$ for a shape factor $S=1.05$ (figure \ref{fig:StabFluctuation}a) and $S=1.03$ (figure \ref{fig:StabFluctuation}b).  The far-field analysis predicts linear stability when $S=1.05$ and instability when $S=1.03$, and this is observed in the numerical solutions with the amplitude of perturbations decaying with increasing distance downstream when $S=1.05$ (figure \ref{fig:StabFluctuation}a), in contrast to the growth in the amplitude of the perturbations that is seen when $S=1.03$ (figure \ref{fig:StabFluctuation}b).

\section{Similarity solutions for the unsteady evolution of plumes following an abrupt change in the source buoyancy flux}
\label{sec:similarity solns}

We examine now the nonlinear evolution of plumes following an abrupt change in the source conditions.  Specifically, we consider solutions of the system of nonlinear evolution equations \eqref{eqn:flux eqns} with $S>25/24$ (so that the steady states are linearly stable) for an initial boundary value problem in which the magnitude of the buoyancy flux at the source is instantaneously changed from $F_0$ to $F_1$ at $t=0$. For $t<0$, the plume is in a steady state given by \eqref{eqn:steady soln} and \eqref{eqn:steady coeffs}.  We calculate the unsteady evolution as the plume adjusts to the new steady state in which the buoyancy flux, $F_0$, in \eqref{eqn:steady soln} and \eqref{eqn:steady coeffs} is replaced by $F_1$.  Thus, the initial conditions are given by \eqref{eqn:steady soln} and \eqref{eqn:steady coeffs}, while for $t>0$ the new  source conditions for the plume are given by $F(0,t)=F_1$ and $Q(0,t)=M(0,t)=0$.

The numerical solutions demonstrate that the adjustment occurs by the  advection of an unsteady `pulse' through the environment (see figures \ref{fig:up}, \ref{fig:downsmall} and \ref{fig:downbig} for examples of  computations), which is modelled by the nonlinear evolution equations \eqref{eqn:flux eqns}.  In this section we draw out the self-similar adjustment that occurs in the dependent variables.

The similarity variable is established through the following scaling analyses, which require all of the terms in governing partial differential equations \eqref{eqn:flux eqns} to be of the same order.  To this end, to balance all the derivative terms we require $M\sim Qz/t$. Turning then to the `source' terms, from (\ref{eqn:flux eqns}a), we have $Q/z\sim M^{1/2}$, while from (\ref{eqn:flux eqns}b) $M/z\sim QF_1/M$, where we have used $F\sim F_1$ as the scale of the buoyancy flux.  Together these yield $z^4\sim F_1t^3$ and we therefore seek similarity solutions to the system in terms of this gearing between the spatial and temporal scales.  We note that the existence of this similarity grouping was identified by \citet{Scase2006a} \citep[see also][]{Scase2008}, although they did not construct the complete similarity solution; indeed for their system with $S=1$ it was not possible to do so because the ill-posedness of the system manifests itself as irreconcilable divergences in the solution (see \S\ref{sec:well posed}).  Instead \citet{Scase2006a} found a separable solution which did not satisfy the boundary conditions, but did capture some of the features found in their numerical computation.  We discuss in appendix \ref{sec:app local sep sim soln} the counterpart to their separable solution in the now regularised system of governing equations.  Before analysing the form of the similarity solutions, we note that there is another reason for anticipating the similarity scaling deduced above: the system is hyperbolic for $S>1$ and all the characteristics are proportional to $W=M/Q\propto F_{1}^{1/3}z^{-1/3}$.  The characteristic curves are of the form $\dd z/\dd t\propto W$, thus also admitting the similarity scaling $z^4\sim F_{1}t^{3}$ and it is through the evolution of the characteristics that the system adjusts to its new state.

We now seek similarity solutions for the fluxes of volume, momentum and buoyancy, which we write in the following form
\begin{equation}
	Q=\frac {6k}{5}\left(\frac{9k}{10S}\right)^{1/3}F_1^{1/3}z^{5/3}\widehat{Q}(\eta),\quad
	M=\left(\frac{9k}{10S}\right)^{2/3}F_1^{2/3}z^{4/3}\widehat{M}(\eta) \quad \text{and} \quad
	F=F_1\widehat{F}(\eta),
	\label{eqn:simvars}
\end{equation}
where the similarity variable is given by
\begin{equation}
	\eta =\frac{6k}{5}\left(\frac{10S}{9k}\right)^{1/3}\frac{z^{4/3}}{F_{1}^{1/3}t},
\end{equation}
and the similarity functions $\widehat{Q}(\eta)$, $\widehat{M}(\eta)$ and $\widehat{F}(\eta)$ are to be determined.  In this form the steady state given by \eqref{eqn:steady soln} corresponds to constant values of $\widehat{Q}$, $\widehat{M}$ and $\widehat{F}$. It is convenient to further substitute $\widehat{Q}=\eta \widehat{q}$, $\widehat{M}=\eta^{2}\widehat{m}$ and $\widehat{F}=\eta^{3}\widehat{f}$ because this simplifies the governing equations, which are now given by
\begin{equation}
	\begin{pmatrix}
		\displaystyle\frac {4}{3} -2\frac{\widehat{q}}{\widehat{m}} & \displaystyle\frac{\widehat{q}^{2}}{\widehat{m}^{2}} & 0 \\[8pt]
		-1 & \displaystyle\frac{4}{3} S & 0 \\[8pt]
		\displaystyle-\frac{\widehat{f}}{\widehat{m}} & \displaystyle\frac{\widehat{f}\widehat{q}}{\widehat{m}^{2}} & \displaystyle\frac {4}{3} -\frac{\widehat{q}}{\widehat{m}}
	\end{pmatrix}
	\eta\fd{}{\eta} \begin{pmatrix} \widehat{q}\\ \widehat{m}\\ \widehat{f} \end{pmatrix} = 
	\begin{pmatrix}
		\displaystyle\frac{5}{3}\widehat{m}^{1/2}-3\widehat{q} \\[8pt]
		\displaystyle\frac{4}{3} \frac{S\widehat{f}\widehat{q}}{\widehat{m}}-4S\widehat{m}+\widehat{q} \\[8pt]
		\displaystyle-4\widehat{f}+\frac{2\widehat{f}\widehat{q}}{\widehat{m}}
	\end{pmatrix}.
	\label{eqn:simsys}
\end{equation}
Symbolically, we write this coupled differential system as $\tensor{D}\eta\dd\bsym{\widehat{q}}/\dd\eta=\bsym{b}$, noting that both the matrix $\tensor{D}$ and the vector $\bsym{b}$ are only functions of the dependent variables $\bsym{\widehat{q}}=(\widehat{q},\widehat{m},\widehat{f})$. We note that the `separable' similarity solutions derived by \citet{Scase2006a} correspond to $\bsym{\widehat{q}}=(\widehat{q},\widehat{m},\widehat{f})=\mbox{constant}$ and these are discussed in appendix \ref{sec:app local sep sim soln}.  For the motion driven by an abrupt change in the magnitude of the buoyancy flux at source, the vital parameter is the ratio of the initial to final buoyancy fluxes at the source given by $\mathcal{F}=F_{0}/F_{1}$ . We note that the initial conditions correspond to 
\begin{equation}
	(\widehat{q},\widehat{m},\widehat{f})=\left(\frac{\mathcal{F}^{1/3}}{\eta},\frac{\mathcal{F}^{2/3}}{\eta^{2}},\frac{\mathcal{F}}{\eta^{3}}\right)
\end{equation}
In terms of the similarity functions, these are enforced in the far-field ($\eta\to\infty$) and correspond to the region that is unaffected by the change of the buoyancy flux at the source.  The source conditions demand that
\begin{equation}
	(\widehat{q},\widehat{m},\widehat{f})\to \left(\frac{1}{\eta},\frac{1}{\eta^2},\frac{1}{\eta^3}\right)\qquad \text{as} \qquad \eta\to 0.
\end{equation}
Constructing the similarity solutions then corresponds to integrating the governing system of ordinary differential equations \eqref{eqn:simsys}, subject to these conditions.

The matrix $\tensor{D}$ is singular when
\begin{equation}
	\left(\frac{4}{3}-\frac{\widehat{q}}{\widehat{m}}\right)\left(\frac{16}{9}-\frac{8S}{3}\frac{\widehat{q}}{\widehat{m}}+\frac{\widehat{q}^{2}}{\widehat{m}^{2}}\right)=0.
\end{equation}
This corresponds to locations where $\widehat{q}/\widehat{m}=4/3$ and $\widehat{q}/\widehat{m}=4\left(S\pm\sqrt{S^2-S}\right)/3$.  These are singular points of the governing system of equations and are of significance because the dependent variables may have discontinuous gradients at these locations.  We show in appendix \ref{sec:app local expansion} how to derive the local behaviour close to the singular points; this is required to initiate numerical integration from these locations.

The similarity solutions may also feature discontinuous solutions, in which case we define the shock position, $z_s(t)$, scaled according to the similarity variables to be given by
\begin{equation}
	z_{s}^{4/3}=\frac{5}{6k}\left(\frac{9k}{10S}\right)^{1/3}F_1^{1/3}t\eta_s
\end{equation}
where $\eta_s$ is a constant.  In this case of discontinuous solutions, the jump conditions relate the dependent variables at $\eta=\eta_{s}^{+}$ to those at $\eta=\eta_{s}^{-}$ and are given by
\begin{equation}
	\left[\widehat{q}-\frac{3\widehat{q}^{2}}{4\widehat{m}}\right]_{-}^{+}=0, \qquad
	\left[S\widehat{m}-\frac{3\widehat{q}}{4}\right]_{-}^{+}=0 \qquad \text{and} \qquad
	\left[f-\frac{3\widehat{q}\widehat{f}}{4\widehat{m}}\right]_{-}^{+}=0.
	\label{eqn:jump}
\end{equation}

Key locations in the similarity solutions are the points at which the dependent variables transition from the new steady state to an unsteady pulse and then from this unsteady pulse to the original steady state.  The family of slowest moving characteristics associated with the new source is given by
\begin{equation}
	\fd{z}{t}=\frac{M}{Q}\left(S-\sqrt{S\left(S-1\right)}\right)=\frac{z\widehat{m}}{t\widehat{q}}\left(S-\sqrt{S\left(S-1\right)}\right).
\end{equation}
The transition between the steady and unsteady portions of the solution must occur at a singular point of $\tensor{D}$ to permit the gradient of the solution to change discontinuously.  Thus in this case $\widehat{q}/\widehat{m}=4\left(S-\sqrt{S\left(S-1\right)}\right)/3$ and so we find that
\begin{equation}
	\fd{z}{t}=\frac{3z}{4t}.
\end{equation}
This implies that the transition point occurs at a constant value of the similarity variables, $\eta=\eta_{c1}$, given by
\begin{equation}
	\eta_{c1}=\frac{4}{3}\left(S-\sqrt{S\left(S-1\right)}\right).
	\label{eqn:etac1}
\end{equation}
Likewise the family of fastest moving characteristics is given by
\begin{equation}
	\fd{z}{t}=\frac{M}{Q}\left(S+\sqrt{S\left(S-1\right)}\right)=\frac{z\widehat{m}}{t\widehat{q}}\left(S+\sqrt{S\left(S-1\right)}\right).
\end{equation}
and so the boundary between the unsteady pulse and the steady solution associated with the original buoyancy flux occurs at a constant value of the similarity variable, $\eta=\eta_{c2}$ given by
\begin{equation}
	\eta_{c2}=\frac{4}{3}\mathcal{F}^{1/3}\left(S+\sqrt{S\left(S-1\right)}\right),
	\label{eqn:etac2}
\end{equation}
provided the motion is due to a decrease in the source strength $(\mathcal{F}>1)$.  If the source strength increases then the characteristics associated with the new release overtake those due to the original source and, as we show below, the motion forms `shocks'.

\subsection{Increase in buoyancy flux}
\label{subsec:increase}

These flows correspond to a new scaled buoyancy flux such that $\mathcal{F}<1$.  For $\eta<\eta_{c1}$, the similarity solution is given by
\begin{equation}
	\left(\widehat{q},\widehat{m},\widehat{f}\right)=\left(\frac{1}{\eta},\frac{1}{\eta^2},\frac{1}{\eta^3}\right),
\end{equation}
where $\eta_{c1}$ is given by \eqref{eqn:etac1} and corresponds to the slowest moving characteristics associated with the new buoyancy flux.  The leading edge of the unsteady solution corresponds to a shock at $\eta=\eta_s$ such that, for $\eta>\eta_s$, the solution is given by the steady state associated with the original buoyancy flux given by \eqref{eqn:steady soln}.

Constructing the similarity solution then requires the computation of the solution for $\eta_{c1}<\eta<\eta_s$, where $\eta_s$ is to be determined as part of that solution.  We note that this domain includes a location where $\widehat{q}/\widehat{m}=4/3$.  Here there is the potential for a contact discontinuity where the reduced gravity is discontinuous or even unbounded, but the volume and momentum fluxes are continuous.  The numerical problem is therefore a boundary-value, ordinary differential equation with an internal critical point, which we solve using a numerical shooting technique. Our method of solution proceeds as follows. First numerically integrate from $\eta=\eta_{c1}$ using the local series expansion given in appendix \ref{sec:app local expansion}.  This expansion provides a solution local to the critical point and entails an undetermined constant, $C_{\alpha}$.  Given a value of $C_\alpha$, the numerical integration can be continued until $\eta=\eta_{m1}$ at which $\widehat{q}/\widehat{m}=4/3$.  The solution is also numerically integrated from the shock at the leading edge.  Given a shock location $\eta_{s}$, the jump conditions \eqref{eqn:jump} provide the conditions at $\eta=\eta_{s}^{-}$ and then the solutions may be numerically integrated until $\eta=\eta_{m2}$ at which $\widehat{q}/\widehat{m}=4/3$.  The constant $C_\alpha$ and the shock location $\eta_{s}$ are then iteratively adjusted until $\eta_{m1}=\eta_{m2}$ and $Q(\eta_{m1})=Q(\eta_{m2})$ (noting that this ensures that the momentum flux is continuous at this location as well), and when these conditions are satisfied, this provides the entire solution.

\begin{figure}
	\begin{center}
	\SetLabels
	\L (0.23*0.02) $B$ \\
	\L (0.5*0.02) $W$ \\
	\L (0.79*0.02) $F$ \\
	\L (0.05*0.55) $z$ \\
	\R (0.12*0.105) $0$ \\
	\R (0.12*0.205) $5$ \\
	\R (0.12*0.305) $10$ \\
	\R (0.12*0.405) $15$ \\
	\R (0.12*0.509) $20$ \\
	\R (0.12*0.609) $25$ \\
	\R (0.12*0.712) $30$ \\
	\R (0.12*0.813) $35$ \\
	\R (0.12*0.914) $40$ \\
	\L (0.125*0.08) $0$ \\
	\L (0.172*0.08) $20$ \\
	\L (0.225*0.08) $40$ \\
	\L (0.280*0.08) $60$ \\
	\L (0.332*0.08) $80$ \\
	\L (0.405*0.08) $0$ \\
	\L (0.504*0.08) $0.5$ \\
	\L (0.61*0.08) $1.0$ \\
	\L (0.68*0.08) $10^{-2}$ \\
	\L (0.787*0.08) $10^{0}$ \\
	\L (0.895*0.08) $10^{2}$ \\
	\endSetLabels
	\strut\AffixLabels{\includegraphics[width=\textwidth,height=4in]{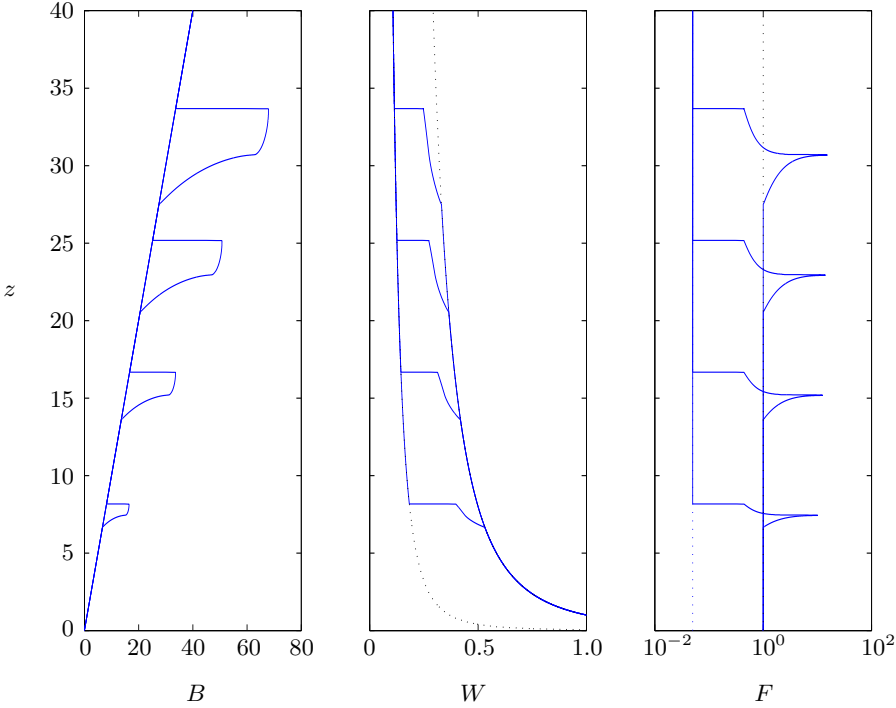}}
	\end{center}
	\caption{ The scaled width, $B(z)=z \widehat{Q}/\widehat{M}^{1/2}$, the scaled velocity, $W(z)=z^{-1/3}\widehat{M}/\widehat{Q}$ and the buoyancy flux, $\widehat{F}$ as functions of the distance from source, $z$ at dimensionless times $t=2.4,4.9,7.4$ and $9.9$ for an increase in buoyancy flux $(\mathcal{F}=0.05)$ at $t=0$ and with shape factor $S=1.1$. In (b) the scaled velocity, $W=z^{-1/3}$, corresponding to the steady velocity field of the original buoyancy flux at the source, and the scaled velocity $W=z^{-1/3}20^{1/3}$, corresponding to the steady velocity field associated with the new buoyancy flux, are plotted in dashed lines.  In (c) the original buoyancy flux, $F=F_{0}=0.05$, and the new steady state buoyancy flux, $F=F_{1}=1$, are plotted in dashed lines.}
	\label{fig:up}
\end{figure}

\begin{figure}
	\begin{center}
	\SetLabels
	\L (0.11*0.73) $\widehat{f}(\eta)$ \\
	\L (0.4*0.73) $\widehat{m}(\eta)$ \\
	\L (0.57*0.73) $\widehat{q}(\eta)$ \\
	\L (0.85*0.83) $\eta=\eta_{s}$ \\
	\L (0.85*0.14) $\eta=\eta_{c1}$ \\
	\L (0.07*0.50) $\eta=\eta_{m}$ \\
	\L (0.036*0.0) $0$ \\
	\L (0.22*0.0) $1$ \\
	\L (0.401*0.0) $2$ \\
	\L (0.583*0.0) $3$ \\
	\L (0.765*0.0) $4$ \\
	\L (0.948*0.0) $5$ \\
	\L (-0.05*0.5) $\eta$ \\
	\L (-0.02*0.04) $1.00$ \\
	\L (-0.02*0.187) $1.05$ \\
	\L (-0.02*0.335) $1.10$ \\
	\L (-0.02*0.485) $1.15$ \\
	\L (-0.02*0.636) $1.20$ \\
	\L (-0.02*0.787) $1.25$ \\
	\L (-0.02*0.938) $1.30$ \\
	\endSetLabels
	\strut\AffixLabels{\includegraphics[width=0.8\textwidth]{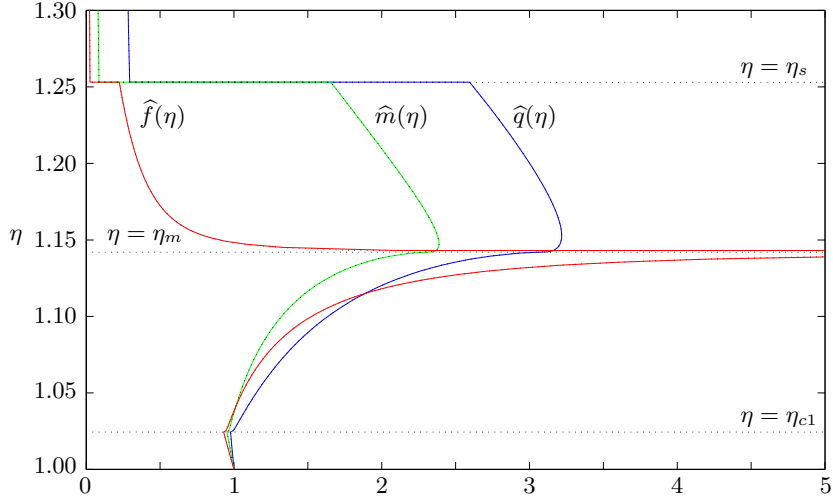}}
	\end{center}
	\caption{The similarity solution for the volume flux, $\widehat{q}(\eta)$, the momentum flux, $\widehat{m}(\eta)$, and the buoyancy flux, $\widehat{f}(\eta)$ as functions of the similarity variable $\eta$ for shape factor $S=1.1$ and an increase in the source buoyancy flux $\mathcal{F}=0.05$ (plotted in solid lines).  Also plotted are results from the direct numerical integration of the governing equations (dashed lines), although the two sets of curves are so close in values that they are virtually indistinguishable.  The values $\eta_{c1}$, $\eta_m$ and $\eta_s$ are also marked.  These corresponds, respectively, to the boundary between the time-dependent part of the solution and the steady state associated with the new buoyancy flux at the source, the location of a contact discontinuity and the location of a shock that marks the interface between the unsteady evolution and the steady steady associated with the original buoyancy flux at the source.}
	\label{fig:upsim}
\end{figure}

The numerical integration of the governing partial differential equations is plotted in figure \ref{fig:up} at various instances of time and the underlying similarity form of solution is plotted in figure \ref{fig:upsim}.  We note that there is excellent correspondence between the two, as evidenced by the overlap between the similarity solution and the direct numerical computations (the curves in figure \ref{fig:upsim} are virtually indistinguishable).  We observe in figure \ref{fig:up} that an increase in buoyancy flux at the source leads to a broadening of the width of the plume in the unsteady pulse before the new steady state is established.  Notably the velocity field is increased relative to the flow associated with the original buoyancy flux.  Together these lead to the surprising variation in the buoyancy flux as it changes from the original value of $\widehat{F}=0.05$ to the new value of $\widehat{F}=1$; the variation is not monotonic and within the unsteady pulse the buoyancy flux overshoots its new value before subsequently decreasing to attain it (see figure \ref{fig:up}c).  The similarity structure is plotted in figure \ref{fig:upsim}, where the similarity variables are plotted between the leading and trailing edges of the unsteady region.  These correspond to a shock at $\eta=\eta_{s}$ and a point where the gradient changes discontinuously at $\eta=\eta_{c1}$.  In between there is a contact discontinuity at $\eta=\eta_{m}$ where the volume and momentum fluxes are continuous, and the buoyancy flux is unbounded.

\subsection{Decrease in buoyancy flux}
\label{subsec:decrease}

When the buoyancy flux decreases relatively weakly $(\mathcal{F}<\mathcal{F}_{m}=6.9)$, we construct the similarity solution between the two critical points, $\eta_{c1}$ and $\eta_{c2}$, given by \eqref{eqn:etac1} and \eqref{eqn:etac2}, respectively.  For $\eta>\eta_{c2}$, the solution corresponds to the steady state associated with the original buoyancy flux, whereas for $\eta<\eta_{c1}$, the solution corresponds to the steady state associated with the new buoyancy flux.  As described above, the boundaries between the steady states and this unsteady pulse are characteristics that propagate at the fastest and slowest rates.  At some point, $\eta_m$, ($\eta_{c1}<\eta_m<\eta_{c2}$) the similarity solution reaches a state in which $\widehat{q}/\widehat{m}=4/3$ and there is a contact discontinuity.

We construct the solutions as follows: we integrate from $\eta=\eta_{c1}$, initiating the numerical solver with the local expansion derived in appendix \ref{sec:app local expansion}, which entrails an adjustable constant, $C_{\alpha 1}$.  We numerically integrate until $\eta=\eta_{m1}$ at which point $\widehat{q}/\widehat{m}=4/3$.  We also numerical integrate from $\eta=\eta_{c2}$, initiating the solution using a local expansion derived in appendix \ref{sec:app local expansion}, which features another adjustable constant $C_{\alpha 2}$.  We integrate until $\eta=\eta_{m2}$ at which point $\widehat{q}/\widehat{m}=4/3$.  The solution is then found by iteratively adjusting $C_{\alpha 1}$ and $C_{\alpha 2}$ until $\eta_{m1}=\eta_{m2}$ and $\widehat{q}(\eta_{m1})=\widehat{q}(\eta_{m2})$.

\begin{figure}
	\begin{center}
	\SetLabels
	\L (0.23*0.02) $B$ \\
	\L (0.5*0.02) $W$ \\
	\L (0.79*0.02) $F$ \\
	\L (0.05*0.55) $z$ \\
	\R (0.12*0.105) $0$ \\
	\R (0.12*0.205) $5$ \\
	\R (0.12*0.305) $10$ \\
	\R (0.12*0.405) $15$ \\
	\R (0.12*0.509) $20$ \\
	\R (0.12*0.609) $25$ \\
	\R (0.12*0.712) $30$ \\
	\R (0.12*0.813) $35$ \\
	\R (0.12*0.914) $40$ \\
	\L (0.125*0.08) $0$ \\
	\L (0.172*0.08) $10$ \\
	\L (0.225*0.08) $20$ \\
	\L (0.280*0.08) $30$ \\
	\L (0.332*0.08) $40$ \\
	\L (0.405*0.08) $0$ \\
	\L (0.504*0.08) $0.5$ \\
	\L (0.61*0.08) $1.0$ \\
	\L (0.68*0.08) $0$ \\
	\L (0.728*0.08) $0.5$ \\
	\L (0.78*0.08) $1.0$ \\
	\L (0.83*0.08) $1.5$ \\
	\L (0.882*0.08) $2.0$ \\
	\endSetLabels
	\strut\AffixLabels{\includegraphics[width=\textwidth,height=4in]{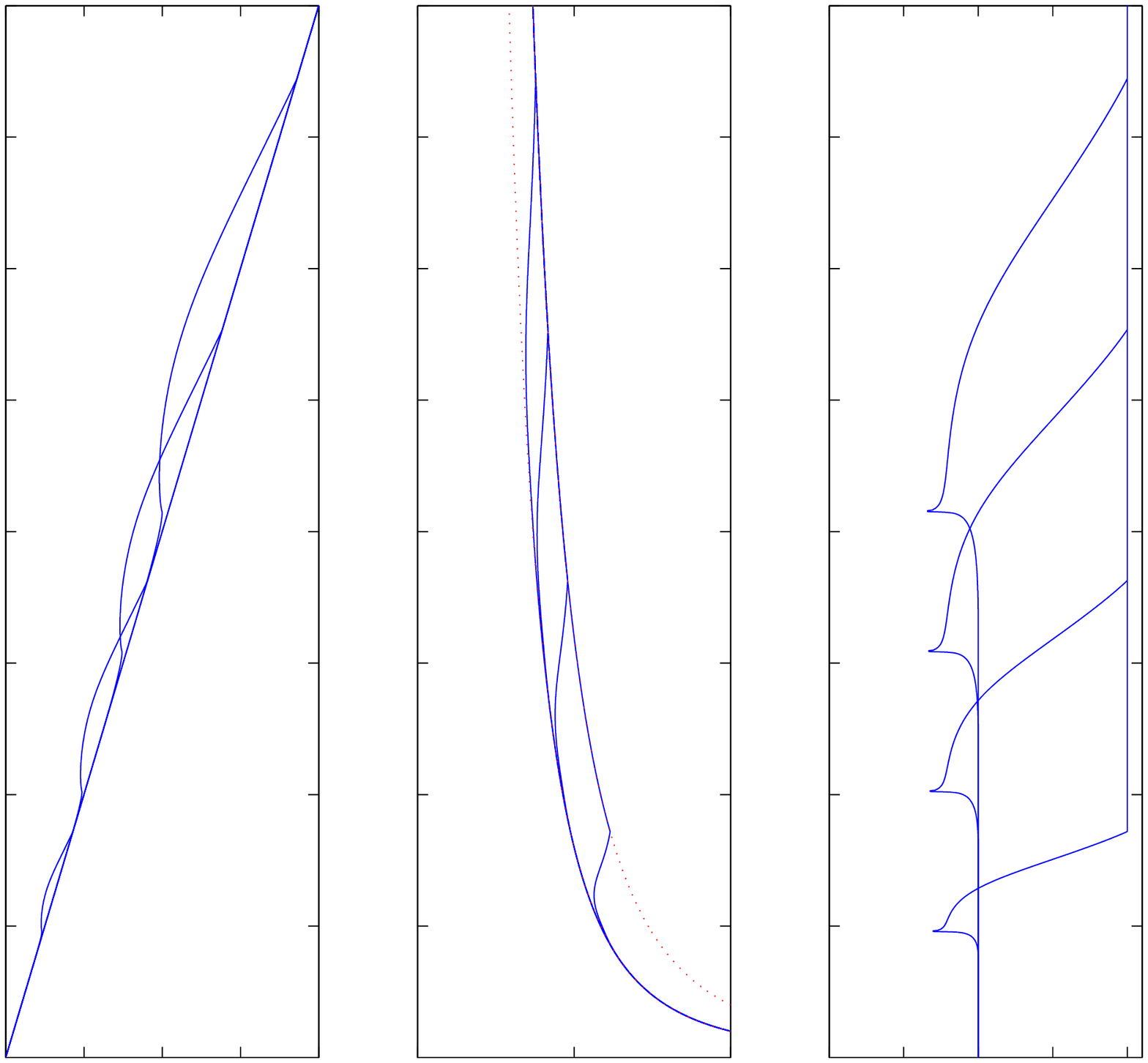}}
	\end{center}
	\caption{ The scaled width, $B(z)=z\widehat{Q}/\sqrt{\widehat{M}}$, the scaled velocity, $W(z)=z^{-1/3}\widehat{M}/\widehat{Q}$ and the buoyancy flux, $F$ as functions of the distance from source, $z$, at dimensionless times $t=4.5,9.5,14.5$ and $19.5$ for a decrease in buoyancy flux $(\mathcal{F}=2)$ at $t=0$ and with shape factor $S=1.1$. In (b) the scaled velocity, $W=z^{-1/3}$, corresponding to the steady velocity field of the original buoyancy flux  at the source, and the scaled velocity $W=z^{-1/3}2^{-1/3}$, corresponding to the steady velocity field associated with the new buoyancy flux, are plotted in dashed lines.}
	\label{fig:downsmall}
\end{figure}

\begin{figure}
	\begin{center}
	\SetLabels
	\L (0.105*0.73) $\widehat{f}(\eta)$ \\
	\L (0.25*0.73) $\widehat{m}(\eta)$ \\
	\L (0.415*0.73) $\widehat{q}(\eta)$ \\
	\L (0.85*0.91) $\eta=\eta_{c2}$ \\
	\L (0.85*0.11) $\eta=\eta_{c1}$ \\
	\L (0.85*0.29) $\eta=\eta_{m}$ \\
	\L (0.036*0.0) $0$ \\
	\L (0.18*0.0) $0.2$ \\
	\L (0.33*0.0) $0.4$ \\
	\L (0.48*0.0) $0.6$ \\
	\L (0.635*0.0) $0.8$ \\
	\L (0.785*0.0) $1.0$ \\
	\L (0.94*0.0) $1.2$ \\
	\L (-0.04*0.5) $\eta$ \\
	\L (0.0*0.055) $1.0$ \\
	\L (0.0*0.354) $1.5$ \\
	\L (0.0*0.645) $2.0$ \\
	\L (0.0*0.938) $2.5$ \\
	\endSetLabels
	\strut\AffixLabels{\includegraphics[width=0.8\textwidth]{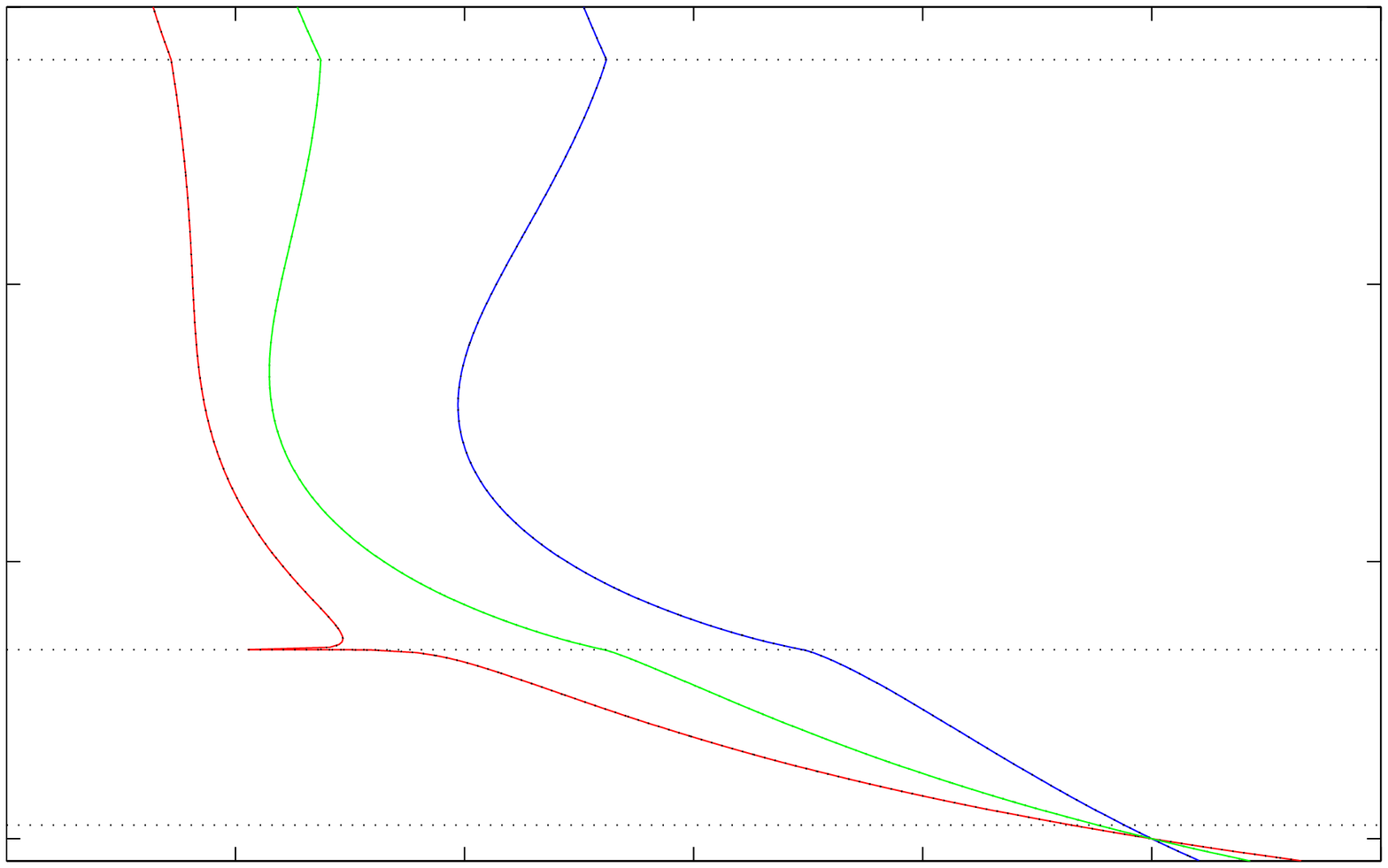}}
	\end{center}
	\caption{The similarity solution for the volume flux, $\widehat{q}(\eta)$, the momentum flux $\widehat{m}(\eta)$ and the buoyancy flux, $\widehat{f}(\eta)$ as functions of the similarity variable $\eta$ for shape factor $S=1.1$ and a decrease in the source buoyancy flux $\mathcal{F}=2$ (plotted in solid lines).  Also plotted are results from the direct numerical integration of the governing equations (dashed lines), although the two sets of curves are so close in values that they are virtually indistinguishable.  The values $\eta_{c1}$, $\eta_m$ and $\eta_{c2}$ are also marked.  These correspond, respectively, to the boundary between the time-dependent part of the solution and the steady state associated with the new buoyancy flux at the source, the location of a contact discontinuity and the location of a the interface between the unsteady evolution and the steady steady associated with the original buoyancy flux at the source.}
	\label{fig:downsmallsim}
\end{figure}

The numerical results from direct integration of the governing partial differential equations are plotted at various instances of time in figure \ref{fig:downsmall}.  Here we observe that the volume and momentum fluxes evolve continuously, in contrast to the evolution following an increase in source strength (\S\ref{subsec:increase}).  During the unsteady evolution from the original state to the new one, the radius of the plume decreases and the velocity increases.  The buoyancy flux, however, does not monotonically vary from the new values $(\widehat{F}=1)$ to its original value $(\widehat{F}=2)$.  Instead it initially decreases (see figure \ref{fig:downsmall}c).  The similarity solution for the unsteady pulse is plotted in figure \ref{fig:downsmallsim} between the leading and trailing characteristics ($\eta_{c1}<\eta<\eta_{c2}$).  This solution features a contact discontinuity at $\eta=\eta_m$, although the mass and momentum fluxes remain continuous at this point.  In figure \ref{fig:downsmallsim} we have plotted both the similarity solution and the rescaled results from the direct numerical integration of the governing equations, and we note that the curves for each of the fields are indistinguishable in this plot.

\begin{figure}
	\begin{center}
	\SetLabels
	\L (0.23*0.02) $B$ \\
	\L (0.5*0.02) $W$ \\
	\L (0.79*0.02) $F$ \\
	\L (0.05*0.55) $z$ \\
	\R (0.12*0.105) $0$ \\
	\R (0.12*0.205) $5$ \\
	\R (0.12*0.305) $10$ \\
	\R (0.12*0.405) $15$ \\
	\R (0.12*0.509) $20$ \\
	\R (0.12*0.609) $25$ \\
	\R (0.12*0.712) $30$ \\
	\R (0.12*0.813) $35$ \\
	\R (0.12*0.914) $40$ \\
	\L (0.125*0.08) $0$ \\
	\L (0.172*0.08) $10$ \\
	\L (0.225*0.08) $20$ \\
	\L (0.280*0.08) $30$ \\
	\L (0.332*0.08) $40$ \\
	\L (0.405*0.08) $0$ \\
	\L (0.450*0.08) $0.5$ \\
	\L (0.504*0.08) $1.0$ \\
	\L (0.557*0.08) $1.5$ \\
	\L (0.61*0.08) $2.0$ \\
	\L (0.68*0.08) $10^{-2}$ \\
	\L (0.787*0.08) $10^{0}$ \\
	\L (0.895*0.08) $10^{2}$ \\
	\endSetLabels
	\strut\AffixLabels{\includegraphics[width=\textwidth,height=4in]{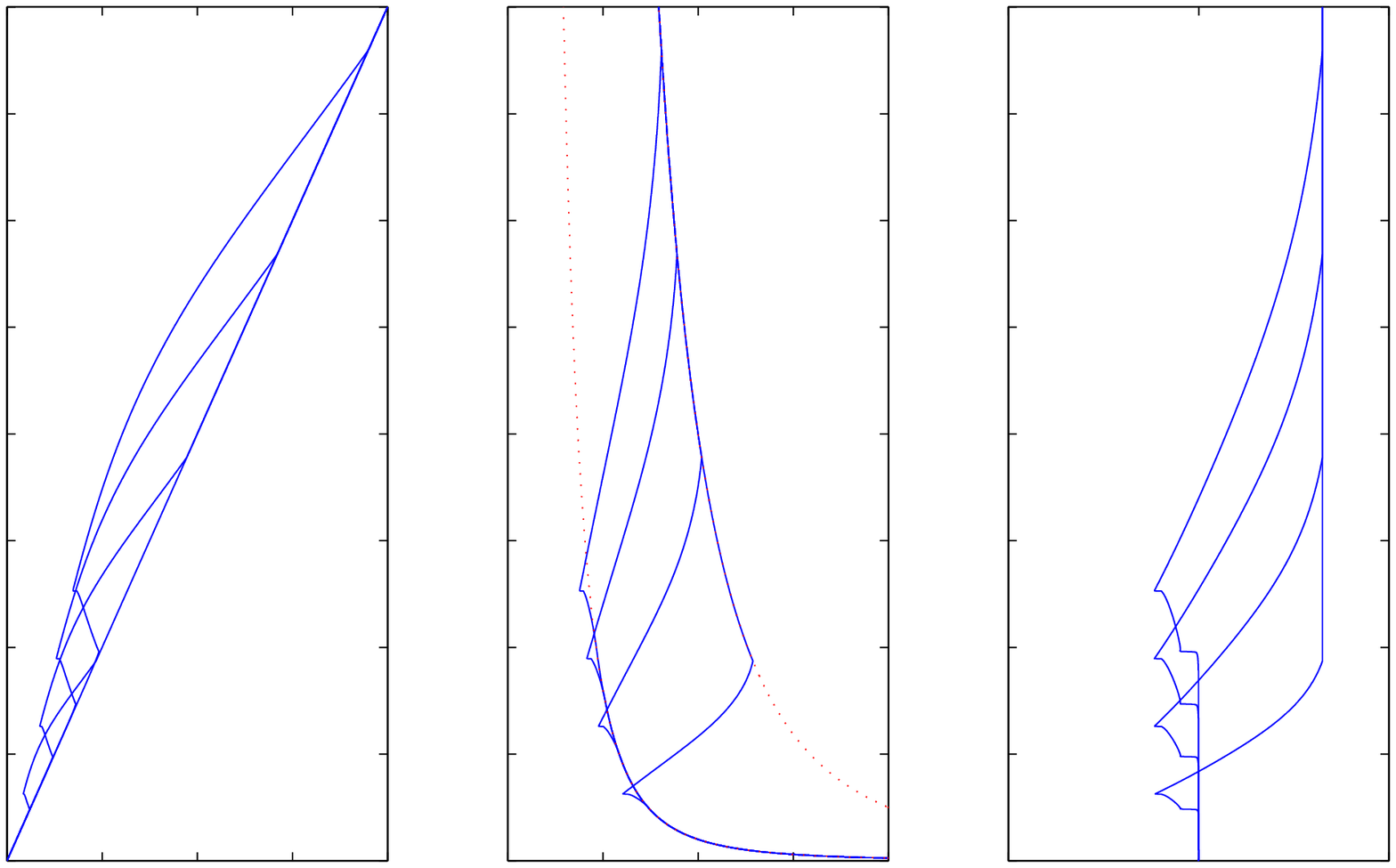}}
	\end{center}
	\caption{ The scaled width, $B(z)=z\widehat{Q}/\sqrt{\widehat{M}}$, the scaled velocity, $W(z)=z^{-1/3}\widehat{M}/\widehat{Q}$ and the buoyancy flux, $F$ as function of the distance from source, $z$, at dimensionless times $t=4.5,9.5,14.5$ and $19.5$ for a decrease in buoyancy flux $\mathcal{F}=0.05$ at $t=0$ and with shape factor $S=1.1$. In (b) the scaled velocity, $W=z^{-1/3}$, corresponding to the steady velocity field of the original buoyancy flux  at the source, and the scaled velocity $W=z^{-1/3}20^{-1/3}$, corresponding to the steady velocity field associated with the new buoyancy flux, are plotted in dashed lines.}
	\label{fig:downbig}
\end{figure}

The morphology of the solution changes for larger decreases in the buoyancy flux.  When $\mathcal{F}>\mathcal{F}_m$, we find that the similarity solution features another critical point at $\eta=\eta_{c3}$ at which $\mathrm{det}(\tensor{D})$ vanishes and the system potentially becomes singular.  Local analysis of the behaviour close to this new critical point indicates that non-integer powers in the series expansion are not possible here; the determined power $\alpha$ is negative and so in order to ensure the fields are bounded, we must enforce $C_{\alpha 3}=0$.  This implies that the dependent variables pass smoothly through this critical point.  However, having attained a state in which $\widehat{q}/\widehat{m}>4(S+\sqrt{S\left(S-1\right)})/3$, the only way to connect to the rest of the solution is via an internal shock at $\eta=\eta_{s}$.  Our method for constructing the solution then proceeds as follows.  We integrate numerically from the critical point at $\eta=\eta_{c1}$, initiating the solution using the local series expansion about this point and introducing an adjustment constant, $C_{\alpha 1}$, to a location $\eta=\eta_{m1}$ at which there is a contact discontinuity ($\widehat{q}/\widehat{m}=4/3$).  We also numerically integrate from $\eta=\eta_{c2}$, initiating the solution using the series expansion with constant $C_{\alpha 2}$.  This constant is adjusted so that an internal critical point is reached at $\eta=\eta_{c3}$ ($\eta_{c3}<\eta_{c2}$), where we enforce the solvability condition given in \eqref{eqn:solvability}.  We may then integrate further; we smoothly pass through the critical point and insert a shock at $\eta=\eta_s<\eta_{c3}$ and then integrate until $\widehat{q}/\widehat{m}=4/3$ at $\eta=\eta_{m2}$.  This leaves two adjustable constants, namely $C_{\alpha 1}$ and $\eta_s$, which are iteratively adjusted until $\eta_{m1}=\eta_{m2}$ and $\widehat{q}(\eta_{m1})=\widehat{q}(\eta_{m2})$ to give the complete solution.

\begin{figure}
	\begin{center}
	\SetLabels
	\L (0.08*0.8) $\widehat{f}(\eta)$ \\
	\L (0.22*0.8) $\widehat{m}(\eta)$ \\
	\L (0.37*0.8) $\widehat{q}(\eta)$ \\
	\L (0.84*0.90) $\eta=\eta_{c2}$ \\
	\L (0.84*0.38) $\eta=\eta_{c3}$ \\
	\L (0.84*0.23) $\eta=\eta_{s}$ \\
	\L (0.84*0.15) $\eta=\eta_{m}$ \\
	\L (0.84*0.08) $\eta=\eta_{c1}$ \\
	\L (0.036*0.0) $0$ \\
	\L (0.155*0.0) $0.2$ \\
	\L (0.285*0.0) $0.4$ \\
	\L (0.416*0.0) $0.6$ \\
	\L (0.546*0.0) $0.8$ \\
	\L (0.676*0.0) $1.0$ \\
	\L (0.805*0.0) $1.2$ \\
	\L (0.94*0.0) $1.4$ \\
	\L (-0.04*0.5) $\eta$ \\
	\L (0.0*0.055) $1.0$ \\
	\L (0.0*0.155) $1.5$ \\
	\L (0.0*0.26) $2.0$ \\
	\L (0.0*0.365) $2.5$ \\
	\L (0.0*0.465) $3.0$ \\
	\L (0.0*0.57) $3.5$ \\
	\L (0.0*0.672) $4.0$ \\
	\L (0.0*0.775) $4.5$ \\
	\L (0.0*0.88) $5.0$ \\
	\endSetLabels
	\strut\AffixLabels{\includegraphics[width=0.8\textwidth]{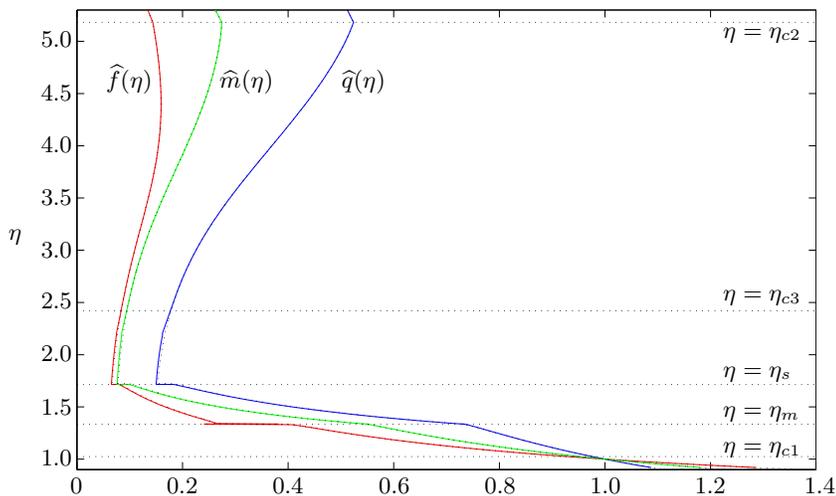}}
	\end{center}
	\caption{The similarity solution for the volume flux, $\widehat{q}(\eta)$, the momentum flux $\widehat{m}(\eta)$ and the buoyancy flux, $\widehat{f}(\eta)$ as functions of the similarity variable $\eta$ for shape factor $S=1.1$ and a decrease in the source buoyancy flux $\mathcal{F}=20$ (plotted in solid lines).  Also plotted are results from the direct numerical integration of the governing equations (dashed lines), although the two sets of curves are so close in values that they are virtually indistinguishable.  The values $\eta_{c1}$, $\eta_m$, $\eta_s$, $\eta_{c3}$ and $\eta_{c2}$ are also marked.  These corresponds, respectively, to the boundary between the time-dependent part of the solution and the steady state associated with the new buoyancy flux at the source, the location of a contact discontinuity, the location of an internal shock between flow states, the location of a continuous transition and the location of a the interface between the unsteady evolution and the steady steady associated with the original buoyancy flux at the source.}
	\label{fig:downbigsim}
\end{figure}

The results from the numerical integration of the governing equations are plotted in figure \ref{fig:downbig} for a large decrease in buoyancy flux at various instances of time.  We note that, as with the weaker decreases in flux, the plume responds by narrowing and accelerating in order to adjust back to its original state.  However a small internal shock is also developed.  For these parameter values ($\mathcal{F}=20$ and $S=1.1$), the shock is of relatively small magnitude and it generates discontinuities in the velocity and width fields, with the reduced gravity remaining continuous.  From this figure it is not possible to observe the presence of the internal critical point $(\eta=\eta_{c3})$ because all of the variables and their derivatives are continuous.  However, it can be confirmed that there is an internal region within which $\widehat{q}/\widehat{m}>4(S+\sqrt{S\left(S-1\right)})/3$.  The associated similarity solution within the unsteady pulse is plotted in figure \ref{fig:downbigsim}.  As with the weaker decrease in buoyancy flux, this plot features continuous transitions at the leading and trailing edges ($\eta=\eta_{c1}$ and $\eta=\eta_{c2}$, respectively) and a contact discontinuity $(\eta=\eta_m)$.  Additionally, there is an internal critical point ($\eta=\eta_{c3}$) and an internal shock ($\eta=\eta_{s}$).  The rescaled results from the numerical integration of the governing equations are overlain on the similarity solutions and again they are virtually indistinguishable, confirming the presence of this similarity solution in the unsteady dynamics of the plume.

\section{Discussion and conclusion}
\label{sec:conclusion}

Steady plume models are applicable on time scales that are long compared to the eddy turnover time that characterises transient turbulent features, and on time scales shorter than the characteristic time for source variations \citep{Scase2006a}.  In many applications, the effect of variations of the source fluxes of mass, momentum or buoyancy on the plume dynamics are of fundamental importance.  Here, unsteady models are essential.  For an example, during volcanic eruptions the source conditions can fluctuate in time due to unsteadiness in the physical processes occurring in the volcanic conduit and at the vent.  End members of the source unsteadiness are the initiation phase (when an eruption first produces material that becomes buoyant and ascents through the atmosphere) and waning phase (when the eruption comes to an end either through a gradual reduction in the rate at which material is erupted or an abrupt cessation of the activity) of the eruption.

The unsteady model of turbulent plumes analysed by \citet{Scase2006a} and \citet{Scase2009} identified a key feature of the unsteady response of plumes to abrupt changes in the source buoyancy flux; the plume adjusts to the new source conditions through a `pulse' that propagates through the plume.  However, the model of \citet{Scase2006a} adopts top-hat profiles for the mean axial velocity and was shown to be ill-posed in the analysis of \citet{Scase2012}, with the numerical solutions of \citet{Scase2006a} and \citet{Scase2009} attained only due to significant numerical viscosity.

The diffusive regularisation of the unsteady model proposed by \citet{Scase2012} has an intuitive physical basis: turbulent eddies have a vertical length scale over which different levels of the plume are connected.  Numerical solutions \citep{Scase2012} suggest that the phenomenological model of the turbulent diffusion of momentum advocated by \citet{Scase2012} results in a well-posed unsteady model, but the analysis presented here shows that this modification of the governing system of equations renders the classical power-law solution for steady plumes unstable.  As the power-law solutions have strong empirical support \citep[e.g.][]{Morton1956,Papanicolaou1988,Shabbir1994}, we conclude that the diffusive regularisation of \citet{Scase2012} is not appropriate.

Our analysis identifies a different physical process, a difference in the rates of transport of the cross-sectionally averaged mass and momentum of the plume (referred to as type I dispersion by \citealt{Craske2015a,Craske2015b}) that occurs due to non-uniform radial profiles of the mean axial velocity in the plume, and we have shown that including this process through a momentum shape factor leads to a well-posed system of equations.  However, this requires the top-hat description of the radial profiles for the axial velocity in the plume, which has been applied extensively (for convenience) in models of steady plumes, to be replaced with a description of the radial variation.  The resulting integral model for unsteady plumes introduces only one additional parameter, the momentum shape factor, over the classical steady plume model \citep{Morton1956}.  Furthermore, the (ensemble averaged) radius of the plume remains dependent on the entrainment coefficient alone, so the unsteady model does not preclude calibration of the entrainment coefficient from laboratory experiments that measure the steady plume width (or spreading angle).  To determine the momentum shape factor, the radial profile of the axial velocity can be measured directly \citep[e.g.][]{Papanicolaou1988,Shabbir1994} and the shape factor computed from \eqref{eqn:S defn}.

Explicitly including a momentum shape factor that differs from unity results in a strictly hyperbolic system of equations that governs the plume dynamics.  This is appealing as laboratory experiments \citep{Scase2008} and numerical modelling \citep{ScaseAspdenCaulfield2009} show that the adjustment of the plume to changes in the source conditions occurs through the propagation of an unsteady pulse, as predicted by our non-diffusive hyperbolic model.  The development of the unsteady pulse following an abrupt change in the source buoyancy flux is described by a similarity solution of the hyperbolic system of equations.  The construction of the similarity solution allows us to identify three qualitatively different regimes of the unsteady evolution.  Following an increase in the source buoyancy flux, the pulse takes the form of a localised increase in the plume width with a leading discontinuity.  If the source buoyancy flux is reduced then the plume width narrows.  For a relatively strong reduction in the source buoyancy flux, an internal shock occurs in the similarity solution, whereas no internal shock is found in the source buoyancy flux is reduced by a smaller amount.  We expect diffusive processes will act to smooth locally the sharp gradients that appear in solutions of the hyperbolic model.  However, hyperbolic models have been shown to capture the dominant flow dynamics in many settings \citep{Whitham}.

\citet{Craske2015a} identify, from direct numerical simulations, the dispersion of momentum as a fundamental feature of turbulent jets and construct an integral model to describe unsteady jets that includes a description of the non-uniform radial profile of the vertical velocity \citep{Craske2015b} through shape factors.  In the model of \citet{Craske2015b} the integral conservation equations are derived from the point-wise momentum and energy conservation equations, following the approach of \citet{PriestleyBall1955}, and an energy shape factor rather than a momentum shape factor is introduced.  The resulting system of integral conservation equations shares some features with the unsteady integral model proposed here (albeit for a jet rather than a plume), in particular the hyperbolic structure of the system of equations.


For jets, the momentum of the flow as it is ejected from a source drives the motion, and there is significant evolution of the radial profile of the axial velocity \citep[referred to as type II dispersion by][]{Craske2015a,Craske2015b} as the flow develops away from the source.  The evolution of the shape of the radial profiles of axial velocity and buoyancy have been incorporated into a non-constant entrainment coefficient by \citet{Kaminski2005} and \citet{Carazzo2006}.  Furthermore, in the unsteady integral model of \citet{Craske2015b}, the deviation of the radial profiles from self-similar forms gives rise to diffusive terms in the system of integral conservation equations.   Our analysis shows that a description of momentum dispersion, through a shape factor that differs from unity, is sufficient to obtain a well-posed model of unsteady plumes.  Therefore, while for jets it is necessary to account for the evolution of the radial profile of axial velocity to self-similar form, for unsteady pure plumes the classical entrainment closure of \citet{Morton1956} remains applicable.

The mathematical model we present allows predictions of unsteady plume dynamics to be made and, while our study has focussed on changes to the source buoyancy flux, the effect of general temporal variations in the source conditions can be examined.  Furthermore, our framework can be applied to describe the unsteady dynamics of plumes in industrial and environmental settings, where additional physical processes such as an ambient flow, thermodynamics and particle transport have a strong influence on the evolution of the plume.  \\

We are grateful to Professor R.R. Kerswell and Dr C.G. Johnson for valuable discussions.  This project has received funding from the European Union's Seventh Programme for research, technological development and demonstration under grant agreement No. 308377 and from the NERC through grant NE/I01554X/1.

\appendix

\section{Conservation of energy for unsteady plumes}
\label{sec:app energy closure}

In the derivation of our integral model for unsteady plumes, we made direct use of the point-wise continuity equation to obtain an integral equation for conservation of mass, following the approach used by \citet{Morton1956} for steady plumes.  This requires \textit{a priori} a choice to be made for a representative plume radius, $b$, and an entrainment closure to describe the mixing of ambient fluid into the plume.

An alternative approach, pioneered by \citet{PriestleyBall1955} for steady flows and developed by \citet{Fox1970}, \citet{Kaminski2005} and \citet{Carazzo2006}, is to substitute conservation of axial kinetic energy for the continuity equation in the set of point-wise conservation equations.  As the equation for conservation of axial kinetic energy is derived as a (axial velocity) moment of the equation for the conservation of axial momentum, rewritten in conservation form through application of the continuity equation, taking the conservation of axial kinetic energy together with the conservation of axial momentum provides no additional information over the set of point-wise conservation equations with continuity and conservation of axial momentum \citep[see][for a detailed discussion]{Morton1971}.  The integral expression for conservation of axial momentum and conservation of axial kinetic energy can be manipulated into a form that expresses conservation of mass in the integral sense and, from this, an expression for turbulent entrainment is obtained \citep{PriestleyBall1955,Fox1970,Morton1971,Kaminski2005,Carazzo2006}.  However, the integral equations that result from the use of the conservation of axial kinetic energy differ from those obtained when the continuity equation is used directly.  Indeed, the differences occur due to the application of turbulent closures in different terms following the cross-sectional integration of the point-wise conservation equations: in the approach of \citet{Morton1956} the turbulence closure is applied as a inflow velocity in the kinematic condition applied at the plume boundary, whereas the closure is applied to the turbulent fluctuation terms that occur in the conservation of axial momentum equation when the approach of \citet{PriestleyBall1955} is used.  Finally, as noted by \citet{Morton1971}, both closures introduce parameters that can only be determined empirically.

Recently, \citet{Craske2015a,Craske2015b} used the conservation of axial kinetic energy is place of the continuity equations in the set of point-wise conservation equations in their model of unsteady jets.  In this appendix we adopt a similar approach for buoyant plumes.  We note that our approach differs as we again specify the plume radius $r=b(z,t)$ as a measurable surface in the plume, although we do not define this precisely, and introduce shape factors to account for non-uniform radial profiles of mean flow quantities.  Thus, we define the volume flux, axial momentum flux, and the flux of mean axial kinetic energy as
\refstepcounter{equation}
$$
	b^{2}W = 2\int_{0}^{b} r \bar{w}\dd r, \quad Sb^{2}W^{2} = 2\int_{0}^{b} r\bar{w}^{2}\dd r, \quad \text{and} \quad S_{e}b^{2}W^{3} = 2\int_{0}^{b} r\bar{w}^{3}\dd r,
	\eqno{(\mathrm{\theequation}{\mathit{a},\mathit{b},\mathit{c}})}
	\label{eqn:integral Q M E}
$$
respectively, where, in (\ref{eqn:integral Q M E}\textit{c}), $S_{e}$ is an energy flux shape factor.

Integration of the point-wise Reynolds-averaged conservation equations for axial momentum \eqref{eqn:RA mom} , axial kinetic energy \citep[see][]{Craske2015a} and reduced gravity \eqref{eqn:RA buoy} gives the following set of integral equations,
\begin{subeqnarray}
	& & \fpd{}{t}\left(b^{2}W\right) + \fpd{}{z}\left(Sb^{2}W^{2}\right) = b^{2}G' - \fpd{}{z}\left(M_{f}+M_{p}\right), \slabel{eqn:int mom Craske} \\[3pt]
	& & \fpd{}{t}\left(Sb^{2}W^{2}\right) + \fpd{}{z}\left(S_{e}b^{2}W^{3}\right) = 2\phi b^{2}G'W \nonumber \\
	& & \qquad + \left(P_{m}+P_{f}+P_{p}\right) - \fpd{}{z}\left(E_{f}+E_{p}\right), \slabel{eqn:int engy Craske} \\[3pt]
	& &\fpd{}{t}\left(b^{2}G'\right) + \fpd{}{z}\left(\phi b^{2}G'W\right) = -\fpd{B_{f}}{z}, \slabel{eqn:int buoy Craske}
	\label{eqn:int Craske}
\end{subeqnarray}
where, following \citet{Craske2015a,Craske2015b}, the correlation fluctuation terms and non-hydrostatic pressure terms, that are formally higher order in the plume slenderness, $R/H$, than the mean fluxes, are retained and given by
\begin{equation}
\left. \begin{array}{l}
\displaystyle
	M_{f} = 2\int_{0}^{b} r\overline{w'^{2}}\dd r, \quad M_{p} = 2\int_{0}^{b} r\bar{p}\dd r, \quad E_{f} = 4\int_{0}^{b} r\overline{w'^{2}}\fpd{\bar{w}}{z}\dd r, \\[16pt]
\displaystyle
	E_{p} = 4\int_{0}^{b}r\bar{p}\bar{w}\dd r, \quad P_{m} = 4\int_{0}^{b} r\overline{u'w'}\fpd{\bar{w}}{r}\dd r, \quad P_{f} = 4\int_{0}^{b} r\overline{w'^{2}}\fpd{\bar{w}}{z}\dd r, \\[16pt]
\displaystyle
	 P_{p} = 4\int_{0}^{b} r\bar{p}\fpd{\bar{w}}{z}\dd r, \quad B_{f} = 2\int_{0}^{b} r\overline{g'_{r}w'}\dd r.
\end{array} \right\}
\label{eqn:fluct integrals}
\end{equation}
Here $\bar{p}$ denotes the averaged deviation of the pressure in the plume from hydrostatic pressure.  Note, in \eqref{eqn:int Craske} we have neglected terms in which flow quantities are evaluated on the plume boundary.

Manipulation of the integral equations for conservation of axial momentum \eqref{eqn:int mom Craske} and conservation of axial kinetic energy \eqref{eqn:int engy Craske} leads to an expression akin to conservation of mass,
\begin{equation}
	\frac{S}{S_{e}}\fpd{}{t}\left(b^{2}\right) + \fpd{}{z}\left(b^{2}W\right) = 2k_{e}bW,
	\label{eqn:int mass from engy}
\end{equation}
where the entrainment rate $k_{e}$ is given by
\begin{eqnarray}
	k_{e} &=& \left(\frac{S-\phi}{S_{e}}\right)\frac{bG'}{W^{2}} + \frac{1}{bW^{2}}\left(\frac{1}{S}-\frac{S}{S_{e}}\right)\fpd{}{z}\left(Sb^{2}W^{2}\right) + \frac{b}{2S_{e}W}\fpd{S}{t} + \frac{S^{2}}{2S_{e}}\fpd{}{z}\left(\frac{S_{e}}{S^{2}}\right) \nonumber \\[3pt]
	&& - \frac{P_{m}+P_{m}+P_{f}}{2S_{e}bW^{3}} - \frac{S}{S_{e}bW^{2}}\fpd{}{z}\left(M_{f}+M_{p}\right) + \frac{1}{2S_{e}bW^{3}}\fpd{}{z}\left(E_{f}+E_{p}\right).
	\label{eqn:entrainment engy}
\end{eqnarray}

The form of the mass conservation equation \eqref{eqn:int mass from engy} is similar to the expression proposed by \citet{Craske2015a,Craske2015b}, but differs due to different choices for the plume width.  The entrainment rate has a similar form to that of \citet{Craske2015a,Craske2015b} for jets, although in \eqref{eqn:entrainment engy} the first term on the right-hand-side does not appear for jets where $G'\equiv 0$.  A similar term, which results in an entrainment rate that depends on the local Richardson number of the plume (given by $\Ri = bG'/W^{2}$), occurs in the entrainment rate in the models that adopt the approach of \citet{PriestleyBall1955} \citep[e.g.][]{PriestleyBall1955,Fox1970,Kaminski2005,Carazzo2006} and has been shown to be important when describing the near source development of buoyant jets into plumes \citep{Kaminski2005,Carazzo2006,Ezzamel2015}.  Further, the entrainment rate here includes a contribution from temporal changes in the momentum shape factor that does not appear in the entrainment rate of \citet{Craske2015a,Craske2015b}.

For turbulent plumes from pure plume sources at distances sufficiently far from the source, experiments suggest the radial profiles of mean axial velocity and reduced gravity, and the second-order turbulent velocity statistics, have reached a self-similar form \citep{Ezzamel2015}.  The shape factors can then be taken as constant values.  Furthermore, following \citet{Craske2015b} and for simplicity, we take the turbulent production and transport terms and non-hydrostatic pressure terms to be such that the entrainment rate is given by,
\begin{equation}
	k_{e} \approx k_{0} + \left(\frac{S-\phi}{S_{e}}\right)\frac{bG'}{W^{2}} + \frac{1}{bW^{2}}\left(\frac{1}{S}-\frac{S}{S_{e}}\right)\fpd{}{z}\left(Sb^{2}W^{2}\right),
	\label{eqn:plume var entrainment}
\end{equation}
with $k_{0}$ constant.

Steady solutions of the plume model with the non-constant entrainment rate \eqref{eqn:plume var entrainment} with pure plume boundary conditions ($Q(0)=0$, $M(0)=0$, $F(0)=F_{0}$) are given by,
\begin{subeqnarray}
	& & Q = \frac{6k_{0}}{5}\left(\frac{9k_{0}}{10}\right)^{1/3}\left(\frac{F_{0}}{S}\right)^{1/3}\left[1-\frac{8S}{5S_{e}}\left(S-S^{2}-\phi+S_{e}\right)\right]^{-4/3}z^{5/3}, \slabel{eqn:steady Q var entr} \\[3pt]
	& & M = \left(\frac{9k_{0}}{10}\right)^{2/3}\left(\frac{F_{0}}{S}\right)^{2/3}\left[1-\frac{8S}{5S_{e}}\left(S-S^{2}-\phi+S_{e}\right)\right]^{-2/3}z^{4/3}, \slabel{eqn:steady M var entr} \\[3pt]
	& & F = F_{0}. \slabel{eqn:steady F var entr}
	\label{eqn:steady soln var entr}
\end{subeqnarray}
Therefore, the non-constant entrainment coefficient does not change the spatial variation of the steady solutions, although the pre-factors in the power-law forms are changed from the \citet{Morton1956} solutions, unless the shape factors take values such that $S^{2}-S-S_{e}+\phi =0$.  We note that the factors could diverge if $1-8S\left(S-S^{2}-\phi+S_{e}\right)/5S_{e} = 0$.  The radius of the plume for the non-constant entrainment coefficient is given by
\begin{equation}
	b = \frac{6k_{0}}{5}\left[1-\frac{8S}{5S_{e}}\left(S-S^{2}-\phi+S_{e}\right)\right]^{-1} z,
\end{equation}
and so the plume radius increases linearly with height as in the model of \citet{Morton1956}.  However, in contrast to the model with a constant entrainment coefficient, the spreading angle of the plume depends on the the shape factors for momentum, buoyancy and energy (unless $S^{2}-S-S_{e}+\phi =0$).

An alternative approach is to augment the system of integral equations \eqref{eqn:int Craske} with the integral expression for conservation of mass \eqref{eqn:int mass}.  We can then interpret the integral expression for conservation of axial kinetic energy as governing the evolution of the momentum shape factor and, in order to fully specify this evolution, a closure is required to describe the evolution of the energy shape factor.  A higher velocity moment of the point-wise conservation equations for mass and momentum would provide an integral equation governing the evolution of the energy shape factor, but would, of course, introduce a new shape factor.  Therefore, a turbulence closure must be invoked at some level within the hierarchy of equations.  Laboratory experiments \citep[e.g.][]{Papanicolaou1988,Wang2002,Ezzamel2015} suggest the momentum shape factor can be taken as a constant value for fully developed turbulent plumes sufficiently far from the source, and this represents a turbulence closure at the level of conservation of mass, momentum and buoyancy.


\section{A phase plane analysis of steady solutions of the plume model with diffusive terms.}
\label{sec:app phase plane}

In \S\ref{sec:Scase problems}, the steady solutions of the integral model for unsteady plumes with phenomenological diffusive terms included to describe turbulent mixing by eddy diffusion were shown to be linearly unstable to small perturbations.  Therefore, while steady solutions of the diffusive system of equations can be found analytically (see \ref{eqn:Scase reg steady solns}) and are modifications of the well-known steady solutions given by \citet{Morton1956}, the solutions cannot be realised.  
However, the linear stability analysis in \S\ref{sec:Scase problems} does not investigate other possible steady states.  In particular, it could be possible that other steady solutions that do not enforce pure plume boundary conditions exist, and that these could be attracting states for the steady diffusive system of equations.

Here we examine the steady plume equations with diffusive terms by treating the system of equations as a dynamical system.  To this end, we allow the volume flux $Q$ to be the independent variable in the system rather than the distance from the source.  We note that, in an unstratified ambient the buoyancy flux is conserved, and so $F\equiv F_{0}$ with $F_{0}$ a specified constant.

We consider first the diffusive term proposed by \citet{Scase2012} (as given in the unsteady momentum conservation equation \ref{eqn:reg mom}).  The equation for the conservation of mass in the steady state allows us to write
\begin{equation}
	\fd{}{z} = 2k M^{1/2}\fd{}{Q},
\end{equation}
and therefore, from the conservation of momentum we find
\begin{equation}
	\kappa Q^{2}\fdd{M}{Q} + \kappa\frac{Q^{2}}{M}\fd{M}{Q}^{2} + \left(3\kappa+S\right)Q\fd{M}{Q} + 2\kappa M = -\frac{F_{0}}{2k}\frac{Q^{2}}{M^{3/2}},
	\label{eqn:Scase M(Q)}
\end{equation}
here treating $M$ as a function of $Q$.  Equation \eqref{eqn:Scase M(Q)} can be further manipulated by introducing new dependent variables
\begin{equation}
	X = Q^{-4/5}M(Q) \quad \text{and} \quad Y = Q^{1/5}\fd{M}{Q}-\frac{4}{5}Q^{-4/5}M(Q),
\end{equation}
and an independent variable $\xi$ with $Q=\exp\left(\xi\right)$ which allows the system to be written as the following autonomous system,
\begin{subeqnarray}
	& & \fd{X}{\xi} = Y, \\[3pt]
	& & \kappa\fd{Y}{\xi} + \kappa\frac{Y^{2}}{X} - \left(\frac{4\kappa}{5}+S\right)Y + \left(\frac{2\kappa}{25}-\frac{4S}{5}\right)X = -\frac{F_{0}}{2k}X^{-3/2}.
	\label{eqn:Scase reg auton sys}
\end{subeqnarray}

\begin{figure}
	\SetLabels
	\L (-0.09*0.45) \begin{sideways} $Q^{1/5}\dd M/\dd Q$ \end{sideways} \\
	\L (0.40*0) $Q^{-4/5}M(Q)$ \\
	\L (0.045*0.07) $0$ \\
	\L (0.145*0.07) $0.5$ \\
	\L (0.253*0.07) $1.0$ \\
	\L (0.367*0.07) $1.5$ \\
	\L (0.478*0.07) $2.0$ \\
	\L (0.590*0.07) $2.5$ \\
	\L (0.702*0.07) $3.0$ \\
	\L (0.813*0.07) $3.5$ \\
	\L (0.925*0.07) $4.0$ \\
	\R (0.04*0.12) $-5$ \\
	\R (0.04*0.39) $0$ \\
	\R (0.04*0.665) $5$ \\
	\R (0.04*0.94) $10$ \\
	\endSetLabels
	\centerline{\strut\AffixLabels{\includegraphics[width=0.6\textwidth,keepaspectratio]{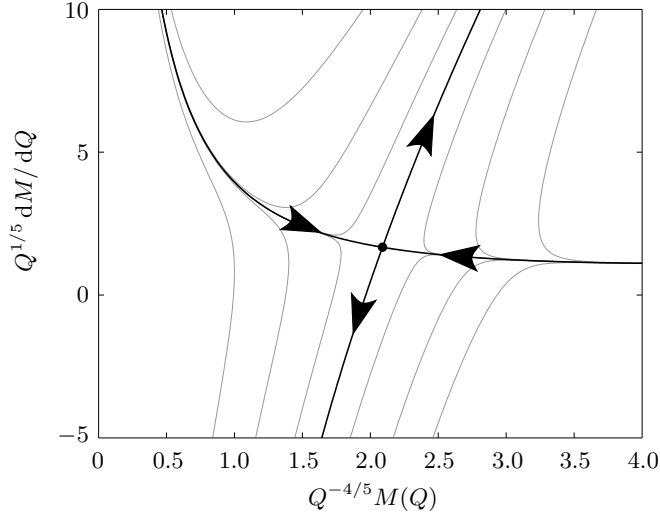}}}
	\caption{Trajectories in the phase-plane of solutions to the steady plume equations with the diffusive regularization of \citet{Scase2012} with $F_{0}=1$, $k=0.1$, $\kappa=0.1$ and $S=1$.  The autonomous system has a single fixed point (denoted by the black point on the figure) that corresponds to the steady pure plume solution of \citet{Morton1956} with a modification for the diffusive term (given by \ref{eqn:Scase reg steady solns}).  The fixed point is a saddle, with two stable directions and two unstable directions, shown as bold solid lines on the figure with arrows denoting the stability of the trajectories.  Additional trajectories, corresponding to solutions with non-pure plume initial conditions, are shown as thin grey lines.}
	\label{fig:RegScaseS1p0PhasePlane}
\end{figure}

\begin{figure}
	\SetLabels
	\L (-0.09*0.45) \begin{sideways} $Q^{1/5}\dd M/\dd Q$ \end{sideways} \\
	\L (0.40*0) $Q^{-4/5}M(Q)$ \\
	\L (0.045*0.07) $0$ \\
	\L (0.145*0.07) $0.5$ \\
	\L (0.253*0.07) $1.0$ \\
	\L (0.367*0.07) $1.5$ \\
	\L (0.478*0.07) $2.0$ \\
	\L (0.590*0.07) $2.5$ \\
	\L (0.702*0.07) $3.0$ \\
	\L (0.813*0.07) $3.5$ \\
	\L (0.925*0.07) $4.0$ \\
	\R (0.04*0.12) $-5$ \\
	\R (0.04*0.39) $0$ \\
	\R (0.04*0.665) $5$ \\
	\R (0.04*0.94) $10$ \\
	\endSetLabels
	\centerline{\strut\AffixLabels{\includegraphics[width=0.6\textwidth,keepaspectratio]{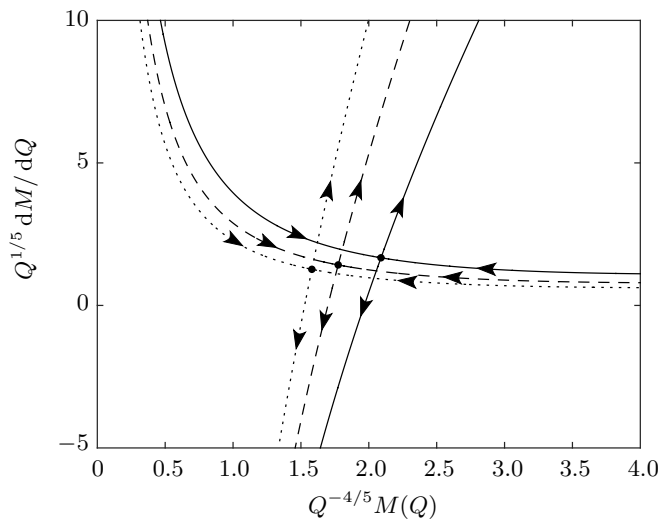}}}
	\caption{The stable and unstable trajectories through fixed points in the phase-plane of solutions to the steady plume equations with the diffusive regularization of \citet{Scase2012} with $F_{0}=1$, $k=0.1$ and $\kappa=0.1$ for momentum shape factor $S=1$ (solid lines), $S=1.5$ (dashed line) and $S=2.0$ (dotted line).}
	\label{fig:RegScaseSPhasePlane}
\end{figure}

The coupled nonlinear system \eqref{eqn:Scase reg auton sys} has a single fixed point at $Y=0$ and $X=X_{0}$ with
\begin{equation}
	X_{0} = \left(\frac{5F_{0}}{8k\left(S-\kappa/10\right)}\right)^{2/5},
\end{equation}
which, after returning to the original variables, is the steady solution given by \eqref{eqn:Scase reg steady solns}.  However, other solutions of the autonomous system for arbitrary initial conditions can be readily found by numerically integrating the system \eqref{eqn:Scase reg auton sys}.  The trajectories in the phase plane corresponding to solutions of \eqref{eqn:Scase reg auton sys} with a momentum shape factor $S=1$ are shown in figure \ref{fig:RegScaseS1p0PhasePlane}.  We see that the fixed point is a saddle with two stable trajectories (such that initial conditions precisely on these trajectories converge to the fixed point) and two unstable trajectories.  For initial conditions that are perturbed away from the pure-plume conditions (i.e. $Q(z=0)=M(z=0)=0$) at the fixed point, the trajectories move the solution away from the fixed point and we find either $M(Q)\to 0$ or $M(Q)\to \infty$ along the phase-plane trajectories.
Taking a momentum shape factor that differs from unity does not change the topology of the trajectories in the phase plane (figure \ref{fig:RegScaseSPhasePlane}).

A local analysis of trajectories that are perturbed away from the fixed point reproduces the results of the linear stability analysis \eqref{eqn:alpha scase}.  Indeed, taking $X=X_{0} + X_{1}(\xi)$ and $Y=Y_{1}(\xi)$ with $\left|X_{1}\right|\ll 1$ and $\left|Y_{1}\right|\ll 1$, we find
\begin{equation}
	\fd{}{\xi}\begin{pmatrix} X_{1} \\ Y_{1} \end{pmatrix} = \begin{pmatrix} 0 & 1 \\  1/5+2S/\kappa & 4/5+S/\kappa \end{pmatrix}\begin{pmatrix} X_{1} \\ Y_{1} \end{pmatrix},
\end{equation}
and the eigenvalues of the Jacobian matrix are
\begin{equation}
	a_{\pm} = \frac{5S + 4\kappa \pm \sqrt{25S^{2}+240S\kappa - 4\kappa^{2}}}{10\kappa}.
\end{equation}
The corresponding eigenvectors are
\begin{equation}
	\bsym{v}_{\pm} = \left(1, a_{\pm}\right)^{\mathrm{T}}.
\end{equation}
Therefore we find $X_{1}\sim \exp{a\xi}$.

\begin{figure}
	\SetLabels
	\L (-0.09*0.45) \begin{sideways} $Q^{1/5}\dd M/\dd Q$ \end{sideways} \\
	\L (0.40*0) $Q^{-4/5}M(Q)$ \\
	\L (0.045*0.07) $0$ \\
	\L (0.145*0.07) $0.5$ \\
	\L (0.253*0.07) $1.0$ \\
	\L (0.367*0.07) $1.5$ \\
	\L (0.478*0.07) $2.0$ \\
	\L (0.590*0.07) $2.5$ \\
	\L (0.702*0.07) $3.0$ \\
	\L (0.813*0.07) $3.5$ \\
	\L (0.925*0.07) $4.0$ \\
	\R (0.04*0.12) $-5$ \\
	\R (0.04*0.39) $0$ \\
	\R (0.04*0.665) $5$ \\
	\R (0.04*0.94) $10$ \\
	\endSetLabels
	\centerline{\strut\AffixLabels{\includegraphics[width=0.6\textwidth,keepaspectratio]{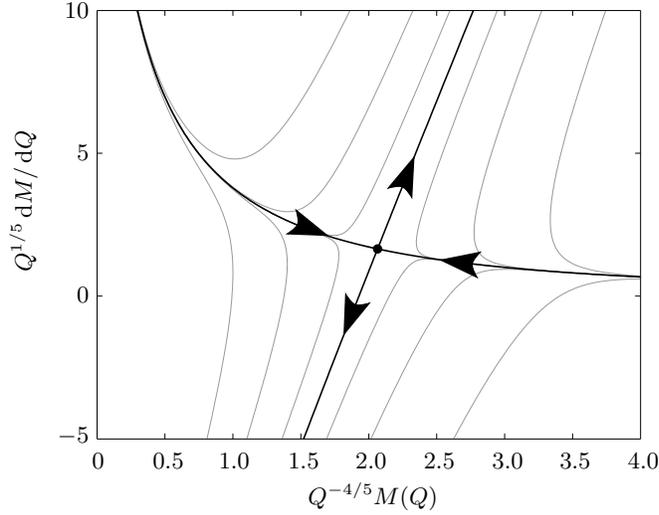}}}
	\caption{Trajectories in the phase-plane of solutions to the steady plume equations with the diffusive regularization \eqref{eqn:reg mom 2} with $F_{0}=1$, $k=0.1$, $\kappa=0.1$ and $S=1$.  The autonomous system has a single fixed point (denoted by the black point on the figure) that corresponds to the steady pure plume solution of \citet{Morton1956} with a modification for the diffusive term (given by \ref{eqn:Scase reg 2 steady solns}).  The fixed point is a saddle, with two stable directions and two unstable directions, shown as bold solid lines on the figure with arrows denoting the stability of the trajectories.  Additional trajectories, corresponding to solutions with non-pure plume initial conditions, are shown as thin grey lines.}
	\label{fig:RegMarkS1p0PhasePlane}
\end{figure}

\begin{figure}
	\SetLabels
	\L (-0.09*0.45) \begin{sideways} $Q^{1/5}\dd M/\dd Q$ \end{sideways} \\
	\L (0.40*0) $Q^{-4/5}M(Q)$ \\
	\L (0.045*0.07) $0$ \\
	\L (0.145*0.07) $0.5$ \\
	\L (0.253*0.07) $1.0$ \\
	\L (0.367*0.07) $1.5$ \\
	\L (0.478*0.07) $2.0$ \\
	\L (0.590*0.07) $2.5$ \\
	\L (0.702*0.07) $3.0$ \\
	\L (0.813*0.07) $3.5$ \\
	\L (0.925*0.07) $4.0$ \\
	\R (0.04*0.12) $-5$ \\
	\R (0.04*0.39) $0$ \\
	\R (0.04*0.665) $5$ \\
	\R (0.04*0.94) $10$ \\
	\endSetLabels
	\centerline{\strut\AffixLabels{\includegraphics[width=0.6\textwidth,keepaspectratio]{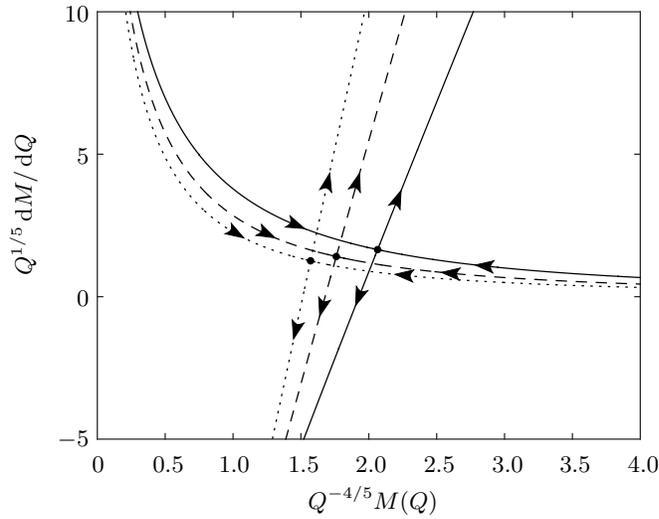}}}
	\caption{The stable and unstable trajectories through fixed points in the phase-plane of solutions to the steady plume equations with the diffusive regularization \eqref{eqn:reg mom 2} with $F_{0}=1$, $k=0.1$ and $\kappa=0.1$ for momentum shape factor $S=1$ (solid lines), $S=1.5$ (dashed line) and $S=2.0$ (dotted line).}
	\label{fig:RegMarkSPhasePlane}
\end{figure}

The analysis presented above can be applied to the alternative form of the eddy diffusion term in the momentum balance given in \eqref{eqn:reg mom 2} for unsteady conditions.  We again find a single fixed point that corresponds to the steady solution \eqref{eqn:Scase reg 2 steady solns}.  The topology of trajectories in the phase plane are qualitatively similar to those of the \citet{Scase2012} diffusive term (figures \ref{fig:RegMarkS1p0PhasePlane} and \ref{fig:RegMarkSPhasePlane}) with the single fixed point being a saddle and trajectories from arbitrary initial conditions have either $M(Q)\to 0$ or $M(Q)\to \infty$ as $Q\to\infty$.  Taking a momentum shape factor that differs from unity does not change the topology of the phase portrait (figure \ref{fig:RegMarkSPhasePlane}).

The conclusion of this analysis is that, for each of the proposed axial diffusion terms (\ref{eqn:Scase diff term} and \ref{eqn:Alt diff term}), the steady solutions for pure plume boundary conditions are not stable and, furthermore, for more general boundary conditions, the steady solutions do not evolve towards the pure plume solution in the far-field.  This is in contrast to steady solutions of the non-diffusive system with general boundary conditions that converge to the pure plume solution in the far-field \citep{Morton1959,Hunt2001,Hunt2005}.


\section{Local series expansions at critical points}
\label{sec:app local expansion}

In this appendix we examine similarity solutions governed by
\begin{equation}
	\tensor{D}\eta \fd{\bsym{\widehat{q}}}{\eta}=\bsym{b},
	\label{eqn:ODE}
\end{equation}
where $\tensor{D}$ and $\bsym{b}$ are the matrix and vector defined in \eqref{eqn:simsys}.  We examine the solutions local to a critical point, $\eta=\eta_{ci}$ ($i=1,2$), of this differential system, where $\mathrm{det}(\tensor{D})$ vanishes.

It is convenient to write $\nu=\log(\eta/\eta_{ci})$ and to write the following expansion of the dependent variables, the matrix $\tensor{D}$ and the forcing vector, $\bsym{b}$,
\begin{subeqnarray}
	\bsym{\widehat{q}}&=&{\bf \widehat{q}}_0+\nu{\bf \widehat{q}}_1+\nu^2{\bf \widehat{q}}_2+\nu^\alpha{\bf \widehat{q}}_\alpha+\nu^{\alpha+1}{\bf \widehat{q}}_{\alpha+1}+\ldots,\slabel{eqn:qseries} \\
	\bsym{\tensor{D}}&=&\tensor{D}_0+\nu \tensor{D}_1+\nu^\alpha \tensor{D}_\alpha+\ldots, \label{eqn:Mseries} \\
	\bsym{b}&=&\bsym{b}_0+\nu\bsym{b}_1+\nu^\alpha\bsym{b}_\alpha+\ldots, \label{eqn:bseries}
	\label{eqn:series}
\end{subeqnarray}
where $\alpha$ is not an integer.  This form of expansion series is possible because $\mathrm{det}(\tensor{D}_0)=0$.  We substitute the series \eqref{eqn:series} into \eqref{eqn:ODE} and in the regime $\nu\ll 1$, balance terms in powers of $\nu$.  At $O(1)$ and at $O(\nu^{\alpha-1})$ we find that
\refstepcounter{equation}
$$
	\tensor{D}_{0}\bsym{\widehat{q}}_1=\bsym{b}_{0} \qquad \text{and} \qquad \alpha \tensor{D}_{0}\bsym{\widehat{q}}_{\alpha}=0.
	\eqno{(\mathrm{\theequation}{\mathit{a},\mathit{b}})}
	\label{eqn:order0}
$$
The matrix $\tensor{D}_0$ is singular; thus we can find vectors $\bsym{e}$ and $\bsym{\hat{e}}^{T}$ such that $\tensor{D}_{0}\bsym{e}=0$ and $\bsym{\hat{e}}^{T}\tensor{D}_0=0$, respectively.  This means that from (\ref{eqn:order0}\textit{a}) we deduce the solvability criterion
\begin{equation}
	\bsym{\hat{e}}^{T}\bsym{b}_0=0,
	\label{eqn:solvability}
\end{equation}
and that $\bsym{\widehat{q}}_{1}=C_{1}\bsym{e}+\bsym{\widehat{q}}_{f}$ and $\bsym{\widehat{q}}_{\alpha}=C_{\alpha}\bsym{e}$, where $\bsym{\widehat{q}}_{f}$ is a particular solution of (\ref{eqn:order0}a) and $C_{1}$ and $C_{\alpha}$ are constants.  From a balance of terms at $O(\nu)$ and $O(\nu^{\alpha})$ we deduce
\refstepcounter{equation}
$$
	\bsym{\hat{e}}^{T}\tensor{D}_{1}\bsym{\widehat{q}}_{1}=\bsym{\hat{e}}^{T}\bsym{b}_{1} \qquad \text{and} \qquad \bsym{\hat{e}}^{T}\left(\tensor{D}_{\alpha}\bsym{\widehat{q}}_{1}+\tensor{D}_{1}\alpha\bsym{\widehat{q}}_{\alpha}\right)=\bsym{\hat{e}}^{T}\bsym{b}_{\alpha}.
	\eqno{(\mathrm{\theequation}{\mathit{a},\mathit{b}})}
	\label{eqn:order1}
$$
The first of these equations (\ref{eqn:order1}\textit{a}) determines the value of the constant $C_{1}$, while the second (\ref{eqn:order1}\textit{b}) determines the non-integer power $\alpha$.

We may apply this formulation to any of the critical points, but the locations that are of most interest are the boundaries between the steady and unsteady portions of the solution. First we analyse the local behaviour near to $\eta_{c2}=4\mathcal{F}^{1/3}\left(S+\sqrt{S\left(S-1\right)}\right)/3$, which corresponds to the boundary between the original source and the unsteady pulse within the similarity solution.  At this point the solution is given by
\begin{equation}
	\bsym{\widehat{q}}_{0}=\left(\beta_{0},\beta_{0}^2,\beta_{0}^3\right), \qquad \text{where} \qquad \beta_{0}=\frac{3}{4}\left(S+\sqrt{S\left(S-1\right)}\right)^{-1}.
\end{equation}
The vectors $\bsym{e}$ and $\bsym{\hat{e}}^{T}$ are given by
\begin{subeqnarray}
	\bsym{e}^{T} &=& \left(\textfrac{16}{9}S\left(2S-1+2\sqrt{S\left(S-1\right)}\right), \textfrac{4}{3}\left(2S-1+2\sqrt{S\left(S-1\right)}\right),1\right), \\[3pt]
	\bsym{\hat{e}}^{T} &=& \left(-\textfrac{3}{4} \left(2S-1-2\sqrt{S\left(S-1\right)}\right),1,0\right).
\end{subeqnarray}
The particular solution is given by
\begin{equation}
	\bsym{\widehat{q}}_{f}=\left(-\beta_{0},-2\beta_{0}^2,-3\beta_{0}^{3}\right),
\end{equation}
the constant $C_{1}=0$ and the exponent, $\alpha$ is 
\begin{equation}
	\alpha = \frac{13}{8} - \frac{5\left(2S-1\right)}{16\sqrt{S\left(S-1\right)}}
\end{equation}
The local expansion is bounded as $\nu\to 0$ provided $\alpha>0$, a condition that demands $S>25/24$. We note that this condition is identical to the criterion for linear stability.  Its origin is identical; it comes from the requirement that the unsteady adjustment remains bounded at its leading edge.

The solution at the trailing edge of the unsteady portion of the solution is broadly similar.  Expanding the dependent variables close to $\eta_{c1}=4\left(S-\sqrt{S\left(S-1\right)}\right)/3$, we find that the solution is given by
\begin{equation}
	\bsym{\widehat{q}}_{0} = \left(\beta_{1},\beta_{1}^{2},\beta_{1}^{3}\right), \qquad \text{where} \qquad \beta_{1} = \textfrac{3}{4}\left(S-\sqrt{S\left(S-1\right)}\right)^{-1}.
\end{equation}
The particular solution is given by
\begin{equation}
	\bsym{\widehat{q}}_{f} = \left(-\beta_{1},-2\beta_{1}^{2},-3\beta_{0}^{3}\right),
\end{equation}
the constant $C_{1}=0$ and the exponent, $\alpha$ is 
\begin{equation}
	\alpha = \frac{13}{8} + \frac{5\left(2S-1\right)}{16\sqrt{S\left(S-1\right)}}.
\end{equation}

\section{Separable similarity solution}
\label{sec:app local sep sim soln}

\citet{Scase2006a} identified a `separable' similarity solution that emerged as an exact solution to their time-dependent governing equations and, although it did not satisfy the boundary conditions exactly, it appeared to play a role in `organising' the underlying dynamics when the buoyancy flux at the origin was abruptly reduced by a relatively large factor.  In this appendix we derive the analogue of their solution in our regularised system of governing equations and discuss its relevance for the solutions that arise when the source buoyancy flux is abruptly changed.

The separable solution emerges as a special case of the similarity solutions derived in \S\ref{sec:similarity solns} and corresponds to a fixed point in the differential system for $\left(\widehat{q}(\eta),\widehat{m}(\eta),\widehat{f}(\eta)\right)$ given by \eqref{eqn:simsys}.  Thus we find that $\bsym{\widehat{q}} = \bsym{\widehat{q}}_{0} \equiv \left(\widehat{q}_0,\widehat{m}_0,\widehat{f}_0\right)$, is given by
\begin{equation}
	\widehat{q}_{0}=\frac{25}{162}, \qquad \widehat{m}_{0}=\frac{25}{324} \qquad \text{and} \qquad \widehat{f}_{0}=\frac{25}{432}\frac{(2S-1)}{S}
\end{equation}
In terms of the original variables this corresponds to
\begin{equation}
	Q=\frac{2k^2z^3}{9t}, \quad
	M=\frac{k^2z^4}{9t^2} \quad \text{and} \quad
	F=\frac{k^2(2S-1)z^4}{9t^3}.
\end{equation}
Notably this solution is independent of the source flux of buoyancy.

\begin{figure}
	\begin{center}
	\SetLabels
	\L (0.25*0.75) $\mathcal{F}=20$ \\
	\L (0.28*0.43) $\mathcal{F}=2$ \\
	\L (0.65*0.3) $\mathcal{F}=0.05$ \\
	\L (0.07*0.5) $\eta$ \\
	\L (0.48*0.0) $\widehat{q}(\eta)/\sqrt{\widehat{m}(\eta)}$ \\
	\R (0.12*0.1) $0$ \\
	\R (0.12*0.235) $1$ \\
	\R (0.12*0.369) $2$ \\
	\R (0.12*0.505) $3$ \\
	\R (0.12*0.638) $4$ \\
	\R (0.12*0.778) $5$ \\
	\R (0.12*0.91) $6$ \\
	\L (0.12*0.07) $0.4$ \\
	\L (0.195*0.07) $0.6$ \\
	\L (0.271*0.07) $0.8$ \\
	\L (0.350*0.07) $1.0$ \\
	\L (0.427*0.07) $1.2$ \\
	\L (0.505*0.07) $1.4$ \\
	\L (0.582*0.07) $1.6$ \\
	\L (0.658*0.07) $1.8$ \\
	\L (0.735*0.07) $2.0$ \\
	\L (0.815*0.07) $2.2$ \\
	\L (0.89*0.07) $2.4$ \\
	\endSetLabels
	\strut\AffixLabels{\includegraphics[width=\columnwidth]{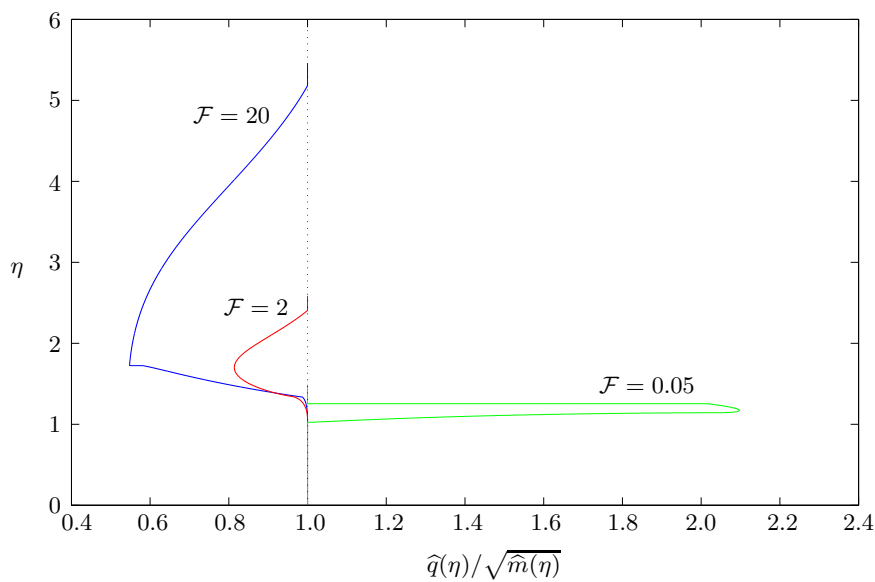}}
	\end{center}
	\caption{The scaled width of the plume, $\widehat{q}/\sqrt{\widehat{m}}$, as a function of the similarity variable, $\eta$, for $\mathcal{F}= 0.05$ (green line, colour online), $\mathcal{F}= 2$ (red line, colour online) and $\mathcal{F}= 20$ (blue line, colour online) and shape factor $S=1.1$.}
	\label{fig:bcomp}
\end{figure}

We may examine the stability of this fixed point in terms of the similarity variable by introducing $\bsym{\widehat{q}}=\bsym{\widehat{q}}_{0}+\bsym{\widehat{q}}_{1}$ and linearizing about the fixed point $\bsym{\widehat{q}}_{0}$.  This gives
\begin{equation}
	\begin{pmatrix}
		-\frac 83 & 4 & 0\\[6pt]
		-1& \frac{4S}{3}& 0\\[6pt]
		-\frac{3(2S-1)}{4} & \frac{3(2S-1)}{2} & -\frac23
	\end{pmatrix}
	\eta \fd{\bsym{\widehat{q}}_{1}}{\eta} =
	\begin{pmatrix}
		-3 & 3 &0\\[6pt]
		\frac{3(2S-1)}{2S}&-\frac{3(2S-1)}{S}& 0\\[6pt]
		2S&2(1-4S)&\frac{8S}{3}
	\end{pmatrix}
	\bsym{\widehat{q}}_{1}.
\end{equation}
We look for a solution of the form $\bsym{\widehat{q}}_{1}= \eta^{\lambda}{\tilde{q}}$ and deduce that
\begin{equation}
	(\lambda+4S)(64S^2\lambda^{2}-72S^2\lambda-72S\lambda^{2}+126S\lambda-162S-36\lambda+81)=0.
\end{equation}
From this condition, we deduce that there is always a root for which $\Real{\lambda}>0$ when $S>1$, while there are other roots for which $\Real{\lambda}<0$. Thus the fixed point, $\bsym{\widehat{q}}_{0}$ is linearly unstable and this implies that in terms of the similarity variables, while the solution may approach the fixed point, it does not remain close to it asymptotically as $\eta \to\infty$.  We illustrate this by plotting $\widehat{q}/\sqrt{\widehat{m}}$ as a function of the similarity variable $\eta$ for $\mathcal{F}=20$, $2$ and $0.05$ (see figure \ref{fig:bcomp}).  A steady state corresponds to $\widehat{q}/\sqrt{\widehat{m}}=1$, while for the separable solution $\widehat{q}/\sqrt{\widehat{m}}=5/9$.  We note that while the evolution for the strongest decrease in buoyancy flux ($\mathcal{F}=20$) becomes close to $\widehat{q}/\sqrt{\widehat{m}}=5/9$ at one location during its evolution, this separable solution does not in general strongly characterise the entire form of the similarity solution.

\bibliography{UnsteadyPlumeBib}
\bibliographystyle{jfm}

\end{document}